\documentclass[prd,twocolumn,nofootinbib,superscriptaddress]{revtex4-2}

\usepackage{preamble}
\usepackage{orcidlink}

\begin{document}
\title{Measuring Supermassive Black Hole Properties via Gravitational Radiation from Eccentrically Orbiting Stellar Mass Black Hole Binaries}
\author{Andrew Laeuger\, \orcidlink{0000-0002-8212-6496}}
\email{alaeuger@caltech.edu}
\affiliation{California Institute of Technology, Pasadena, CA 91126}
\author{Brian Seymour\, \orcidlink{0000-0002-7865-1052}}
\affiliation{California Institute of Technology, Pasadena, CA 91126}
\author{Yanbei Chen\, \orcidlink{0000-0002-9730-9463}}
\affiliation{California Institute of Technology, Pasadena, CA 91126}
\author{Hang Yu\, \orcidlink{0000-0002-6011-6190}}
\affiliation{Montana State University, Bozeman, MT 59717}
\date{September 2023}

\begin{abstract}
    There may exist stellar-mass binary black holes (BBH) which merge while orbiting nearby a supermassive black hole (SMBH).
    In such a triple system, the SMBH will modulate the gravitational waveform of the BBH through orbital Doppler shift and de Sitter precession of the angular momentum.
    Future space-based GW observatories focused on the milli-- and decihertz band will be uniquely poised to observe these waveform modulations, as the GW frequency from stellar-mass BBHs varies slowly in this band while modulation effects accumulate.
    In this work, we apply the Fisher information matrix formalism to estimate how well space-borne GW detectors can measure properties of BBH+SMBH hierarchical triples using the GW from orbiting BBH.
    We extend previous work by considering the more realistic case of an eccentric orbit around the SMBH, and notably include the effects of orbital pericenter precession.
    We find that for detector concepts such as LISA, B-DECIGO, and TianGO, we can extract the SMBH mass and semimajor axis of the orbit with a fractional uncertainty below the 0.1\% level over a wide range of triple system parameters.
    Furthermore, we find that the effects of pericenter precession and orbital eccentricity significantly improve our ability to measure this system.
    We also find that while LISA could measure these systems, the decihertz detector concepts B-DECIGO and TianGO would enable better sensitivity to the triple's parameters.
\end{abstract}

\maketitle
\section{Introduction}
Since the first detection of gravitational waves (GWs), GW astronomy by ground-based detectors has cemented itself as an advantageous method for studying binary systems of compact objects, the majority of which are binary black holes (BBHs) \cite{gwtc1, LIGOScientific:2020ibl, gwtc3}. 
Within the population of observed BBHs, there are systems with progenitors whose masses exceed the predictions of stellar evolution \cite{LIGOScientific:2020iuh,Woosley:2016hmi,Belczynski:2016jno,Spera:2017fyx}. One possible explanation of this detection could be that the progenitors themselves were themselves products of previous mergers \cite{LIGOScientific:2020ufj,Gerosa:2021mno,Antonini:2018auk,2020ApJ...903...67M}. 
The deep potential wells created by supermassive black holes (SMBHs) and their host galactic nuclei could trap the products of stellar mass BBH mergers, making galactic nuclei ideal locations for generating many repeated compact object mergers \cite{PhysRevLett.123.181101,Antonini:2018auk,2020ApJ...903...67M}. Numerical simulations of BBH formation in galactic nuclei due to gas friction \cite{Yang:2020lhq,Bartos:2016dgn} and dynamic capture through gravitational interactions \cite{OLeary:2008myb} suggest that the cosmological merger rate of BBH near galactic nuclei could be of order $\sim$ a few $\text{Gpc}^{-3}\text{yr}^{-1}$. Studying the properties of these repeated merger systems and of the SMBHs which encourage their formation could open a new window on understanding the dynamics of galactic nuclei and the processes which drive galaxy evolution.
The most recent analysis of the BBH population in GWTC-3 is consistent with contributions from both isolated and AGN formations~\cite{KAGRA:2021duu}, though more observations are needed.

In a hierarchical triple system consisting of a stellar-mass BBH orbiting an SMBH, as depicted in Fig.~\ref{fig:Triple system geometry}, the presence of the SMBH would modulate the BBH GW signal through many effects. For example, the velocity of the BBH in its orbit will produce a Doppler shift in the waveform \cite{Randall:2018lnh,Inayoshi:2017hgw,Cardoso:2021vjq}. Allowing the BBH to take an \textit{eccentric} orbit around the SMBH introduces relativistic effects such as pericenter precession as the outer orbital path approaches near the SMBH \cite{Lim:2020cvm}.
Furthermore, the presence of the SMBH will cause the orbital angular momentum of the inner binary $\hat L_i$ to experience de Sitter precession about the orbital angular momentum of the outer binary $\hat L_o$ \cite{Apostolatos:1994mx}. This effect modulates the inclination angle of the BBH angular momentum relative to an observatory in the Solar System. 
The Lidov-Kozai and Lense-Thirring effects also play a role in the evolution of hierarchical triples \cite{Liu:2019tqr,Liu:2021uam}. 

By measuring the effects of Doppler shifts, pericenter precession, and de Sitter precession on the stellar-mass BBH gravitational waveform, one can measure the properties of this triple system, including the SMBH mass, semimajor axis of the outer orbit, and various angles describing the system geometry \cite{Yu:2020dlm,Liu:2021uam,Cardoso:2021vjq,Randall:2018lnh,Inayoshi:2017hgw,Lim:2020cvm,Liu:2019tqr}. These effects accumulate substantially over time scales roughly on the order of an orbital period, which for typical BBH+SMBH triple systems can range from months to years. But because the current ground detectors in LIGO/Virgo/KAGRA are most sensitive between 10 Hz to a few kHz, which correspond to only the final seconds before merger for a stellar-mass BBH, current GW observatories are not optimal for extracting hierarchical triple system parameters through the influence of the SMBH on the waveform \cite{Romero-Shaw:2022ppk,Meiron:2016ipr}.

However, the coming decades could see the construction of a number of proposed space-based detectors which would be sensitive to frequencies below $\sim$1-10 Hz. Building low frequency detectors in space is necessary due to technical challenges from seismic noise \cite{Hall:2020dps, Harms:2013raa} and the need to create arms which are large compared to the curvature of the Earth. The LISA \cite{LISA:2017pwj}, TianQin \cite{TianQin:2015yph}, and Taiji \cite{Hu:2017mde, Ruan:2018tsw} detectors will target the millihertz GW band, while detector concepts such as B-DECIGO \cite{Kawamura:2020pcg, Sato:2017dkf} and TianGO \cite{kunsthesis, Kuns:2019upi} will focus on the decihertz band.
Since the instantaneous orbital decay timescale due to GW emission during inspiral scales roughly with $\omega_{orb}^{-8/3}$ \cite{Maggiore:1900zz}, space-based low-frequency detectors could observe stellar mass BBH for much longer times than ground detectors, making them more favorable for measurements of SMBH-driven effects in the BBH waveform.

Measuring a SMBH with an orbiting binary's GW would be be useful for studying the environment at the center of galaxies. In a recent work by Yu and Chen, it is shown that these proposed low-frequency GW observatories could feasibly measure properties of interest to the few percent level over a wide range of possible BBH+SMBH systems \cite{Yu:2020dlm}. Current observational methods for measuring properties of SMBHs and their local environments include tracking the orbital dynamics of nearby test masses, like stars, and reverberation mapping of the emission line fluxes from the accretion disk, if the SMBH is active \cite{smbhmass}. Recent advances in observational technology and modeling active galactic nuclei have enabled constraints of the masses of their central SMBHs to roughly 10\% precision \cite{Barth2001,Onken2014,Bentz2021,Bentz2022}, though the results obtained by each method do not always agree \cite{EventHorizonTelescope:2019ggy}. Adding a GW-based technique to this toolkit could expand the set of observable SMBHs with well-constrained properties to those which may have few electromagnetic radiation sources nearby \cite{Yu:2020dlm} or foster improvements in established electromagnetic techniques through comparisons of joint measurements. Indeed, there has been significant progress in understanding how space-based GW observatories may be able to measure properties of SMBHs and the objects orbiting them through a variety of triple system phenomena \cite{Yu:2021dqx,Liu:2021uam,Sberna:2022qbn,Kuntz:2022juv}.

The initial work of Yu and Chen assumes a circular Newtonian outer orbit in the BBH+SMBH triple system \cite{Yu:2020dlm}; however, it is expected that formation channels for these systems, especially those which are dynamical in nature, should produce a sizeable population of triples with eccentric outer orbits \cite{Heggie:1975tg}. In this work, we examine how adding a nonzero eccentricity to the outer orbit affects parameter measurement uncertainties. We demonstrate that a nonzero outer eccentricity can significantly improve these uncertainties compared to the circular case, primarily through the inclusion of outer orbit pericenter precession. In order to estimate parameter uncertainties, we rely on the Fisher information matrix, a method which has been frequently used in the past to gauge the measurability of compact binary parameters by ground-based GW observatories \cite{Vallisneri:2007ev}. In short, we find that uncertainties in triple system parameters can consistently fall below the $0.1\%$ level, and that these parameters are measured more precisely with larger $e_o$ and by detectors targeting the decihertz band. We also find that the general trends in parameter measurement are influenced almost entirely by pericenter and de Sitter precession.

In Sec. \ref{sec:Geometry+Waveform}, we outline the mathematical description of the gravitational waveform emitted from a BBH in a hierarchical triple and detected by a space-borne observatory. In Sec. \ref{sec: simplifications}, we outline the Fisher matrix calculation as applied to parameter estimation and explain some simplifications we make to the computation. In Sec. \ref{sec: results}, we present the results of our Fisher matrix computations, and in Sec. \ref{sec: conclusions}, we offer conclusions and possible directions for this work to proceed in the future. In this work, we use geometrized units $G=c=1$.

\section{Mathematical Description of the SMBH+BBH Triple System} \label{sec:Geometry+Waveform}

\subsection{Geometry}

\begin{figure}
    \centering
    \includegraphics[width=\linewidth]{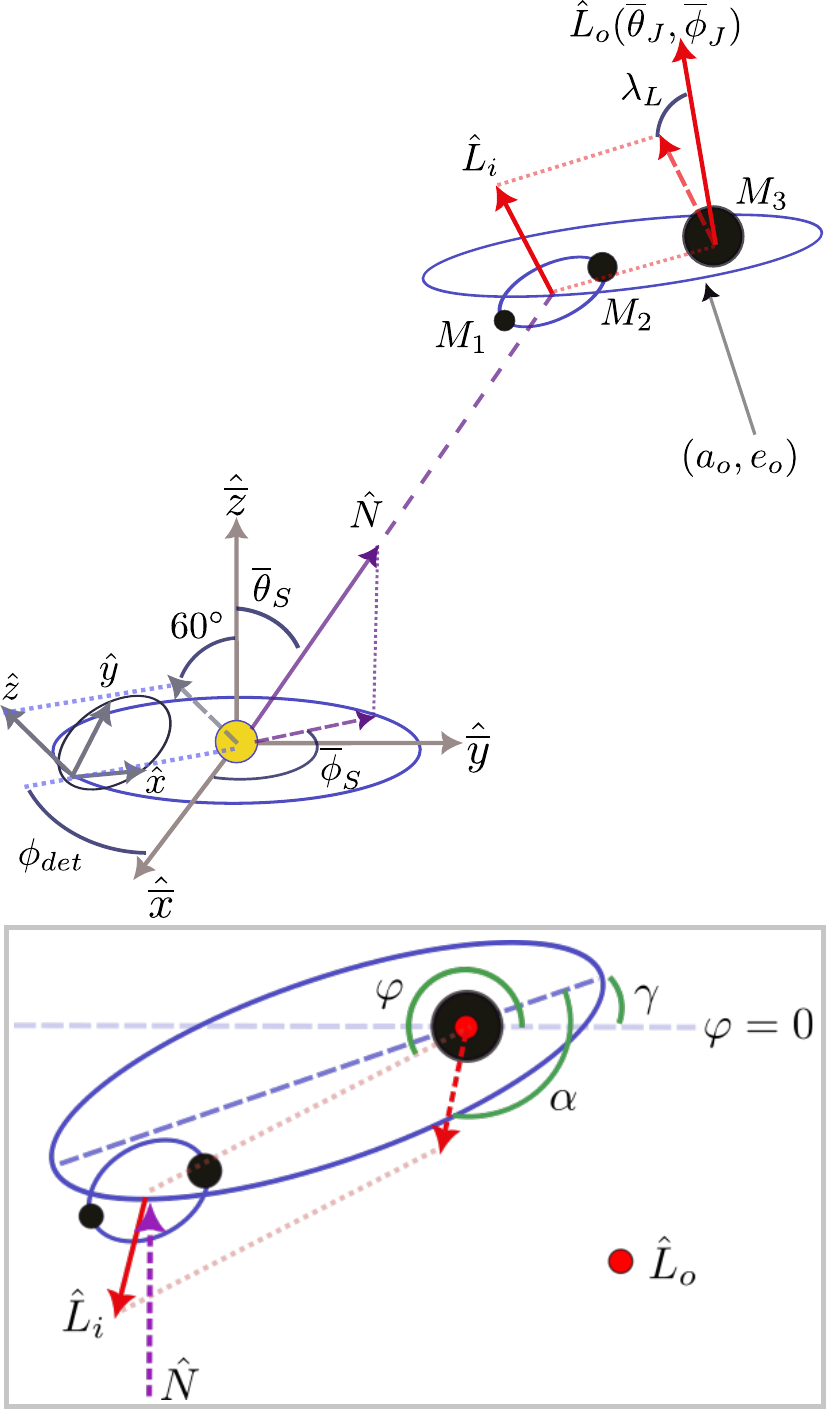}
    \caption{Top: Geometry of the SMBH+BBH triple system. Bottom, inset: View of the triple system normal to the plane of the outer orbit. The outer orbit angular momentum $\hat{L}_o$ points out of the page. See the discussion below and Table \ref{tab:params_table} for definition of all parameters. Figure dimensions are not an indication of true scale.}
    \label{fig:Triple system geometry}
\end{figure}

We first describe the full geometry of the SMBH+BBH triple system with an eccentric outer orbit. Table \ref{tab:params_table} below outlines the set of relevant parameters used in calculating the waveform measured by a space-borne GW observatory. In Fig. \ref{fig:Triple system geometry}, the barred coordinates demarcate a Solar System centered coordinate system, while the unbarred coordinates demarcate a coordinate system based on the orientation of the observatory. 

\begin{table}
    \centering
    \begin{tabular}{|c|c|}
        \hline
        $\boldsymbol{\theta^a}$ & \textbf{Definition }\\
        \hline
        $\log \mathcal{M}_{z}$ &  Detector Frame Chirp Mass: $\mu^{3/5}(m_1+m_2)^{2/5}$ \\
        \hline
        $q$ & Mass Ratio $M_2/M_1$\\
        \hline
        $\log D_L$ & Luminosity Distance \\
        \hline
        $t_c$ & Coalescence Time\\
        \hline
        $\phi_c$ & Coalescence Phase\\
        \hline
        $\overline{\theta}_S$,$\overline{\phi}_S$ & Line of Sight of BBH+SMBH Triple\\
        \hline
        $\overline{\theta}_J$,$\overline{\phi}_J$ & Orientation of Total Angular Momentum $\boldsymbol{J}$\\
        \hline
        $\lambda_L$ & Angle Between $\boldsymbol{L}_i$ and $\boldsymbol{L}_o$\\
        \hline
        $\alpha_0$ & Initial Phase of $\boldsymbol{L}_i$ Around $\boldsymbol{L}_o$\\
        \hline
        $\log M_3$ & SMBH Mass\\
        \hline
        $\log a_o$ & Outer Orbit Semimajor Axis\\
        \hline
        $\gamma_o$ & 
        Initial Outer Orbit Argument of Pericenter (See Note \ref{gamma_o footnote})
        \\
        \hline
        $e_o$ & Outer Orbit Eccentricity\\
        \hline
        $\varphi_0$ & Initial BBH Azimuthal Coordinate\\
        \hline
    \end{tabular}
    \caption{Relevant parameters in BBH+SMBH triple system for GW observed by detectors. Bars over angles indicate the Solar System coordinate frame.
    }
    \label{tab:params_table}
\end{table}
In order to compute the antenna response, we need to be able to convert from the unbarred coordinates to the barred coordinates, which for a constellation-preserving observatory such as LISA, is as follows \cite{Martens:2021phh}:
\begin{gather}
    \hat{x}=-\frac14\sin(2\phi_d)\hat{\overline{x}}+\frac{3+\cos(2\phi_d)}{4}\hat{\overline{y}}+\frac{\sqrt3}2\sin(\phi_d)\hat{\overline{z}}\\
    \hat{y}=\frac{-3+\cos(2\phi_d)}{4}\hat{\overline{x}}+\frac14\sin(2\phi_d)\hat{\overline{y}}-\frac{\sqrt3}2\cos(\phi_d)\hat{\overline{z}}\\
    \hat{z}=-\frac{\sqrt3}{2}\cos(\phi_d)\hat{\overline{x}}-\frac{\sqrt3}{2}\sin(\phi_d)\hat{\overline{y}}+\frac12\hat{\overline{z}}.
\end{gather}
We note that even though B-DECIGO will posses a different detector geometry than LISA
during its orbit, we use the same configuration to simplify the analysis.
The sky location of the hierarchical triple is $(\overline{\theta}_S,\overline{\phi}_S)$, which points along the vector $\hat{N}$, and has a luminosity distance of $D_L$. The triple itself consists of a BBH with black holes of masses $M_1$ and $M_2$, or equivalently, a chirp mass of $\mathcal{M}=\frac{(M_1M_2)^{3/5}}{(M_1+M_2)^{1/5}}$ and mass ratio of $q=M_2/M_1$, and an SMBH of mass $M_3$. The shape of the BBH's orbit around the SMBH can be determined by the semimajor axis $a_o$, the eccentricity $e_o$, 
the angle $\gamma_o$, analogous to the initial Keplerian argument of pericenter\footnote{Of course, the outer orbit is not strictly Keplerian. A rigorous definition of the instantaneous argument of pericenter is subtle, though the picture of an elliptical orbital path with a pericenter that rotates in space at the 1PN-accurate angular velocity of $\frac{3M_3}{a_o(1-e_o^2)}$ is appropriate as a rough approximation. Within the mathematical framework of \cite{cutler_kennefick_poisson_1994}, $\gamma_o$ is implemented as a simple arbitrary rotation of the orbital plane, as in Eq. \eqref{varphi vs chi}. \label{gamma_o footnote}},
and the initial BBH azimuthal coordinate $\varphi_o$.

The unit vector of the angular momentum of the two lighter black holes in the binary system is $\hat{L}_i$, and the unit vector of the angular momentum of the binary's orbit about the SMBH is $\hat{L}_o$. The opening angle $\lambda_L$ is defined by
\begin{equation}
    \cos\lambda_L = \hat{L}_o\cdot\hat{L}_i.
\end{equation}
For $|\vec L_o|>>|\vec L_i|$ and neglecting long time scale orbital effects as well as the spin of the SMBH (see Sec. \ref{subsec:neglected orbital dynamics}), the opening angle stays constant in time, but the orientation of $\hat{L}_i$ traces a cone around $\hat{L}_o$ due to de Sitter precession, with 
\begin{equation}
    \frac{d\hat{L}_i}{dt}=\Omega_{dS}\hat{L}_o\times\hat{L}_i.
\end{equation}
Based on Eq.~(9.200) of \cite{poisson_will_2014}, we use the instantaneous de Sitter precession frequency\footnote{Eq.~(1) of the previous work \cite{Yu:2020dlm} gave the orbit-averaged de Sitter precession rate, which agrees with Eq.~\eqref{eq:Omega_dS}.}
\begin{equation}
    \Omega_{dS}(t)=\frac32\frac{M_3}{r(t)}\dot{\varphi}(t),
    \label{eq:Omega_dS}
\end{equation}
where $r$ is the distance from the SMBH to the center of the BBH and and $\varphi(t)$ is the azimuthal coordinate of the BBH in its orbit (as shown in the inset of Fig. \ref{fig:Triple system geometry}). The orbit-averaged precession rate is 
\begin{equation}
\label{eq:Omega_dS Orbit Averaged}
    \langle\Omega_{dS}\rangle=\frac{3}{2}\frac{M_3}{a_o(1-e_o^2)}\Omega_o,
\end{equation}
where $\Omega_o\equiv\sqrt{M_3/a_o^3}$ is the Newtonian orbital frequency.
The phase of $\hat{L}_i$ in this cone, as shown in the inset of Fig. \ref{fig:Triple system geometry}, can be found by integrating the time-dependent de Sitter precession rate:
\begin{equation}
\label{eq: alpha(t)}
    \alpha(t)=\alpha_0+\int_t^{t_c}\Omega_{dS}(t') dt',
\end{equation}
where $\alpha_0$ is the phase at the time of the binary coalescence $t_c$.

It is also useful to define the inclination angle $\iota_J$ of the outer orbit angular momentum, given by
\begin{equation}
    \cos\iota_J=\hat{N}\cdot\hat{L}_o.
\end{equation}

\subsection{BBH Orbit in Schwarzschild Spacetime}
\label{sec:bbh orbit}
Despite the fact that there does not exist an analytic description of an elliptical orbit in Schwarzschild spacetime, there are well-established methods for computing Schwarzschild geodesics which can be applied to numerically calculate the BBH orbital trajectory~\cite{cutler_kennefick_poisson_1994,1930JaJAG...8...67H, 1959RSPSA.249..180D, 1961RSPSA.263...39D, Chandrasekhar:1985kt}. In particular, we follow the procedure of \cite{cutler_kennefick_poisson_1994}. Defining $p=\frac{a_o}{M_3}(1-e_o^2)$
for semimajor axis $a_o$ and $e_o$, we find a minimum and maximum orbital radius
\begin{equation}
    r_\text{min}=\frac{pM_3}{1+e_o},\quad r_\text{max}=\frac{pM_3}{1-e_o}
\end{equation}
Stable orbits only exist for $p>6 + 2e_o$~\cite{cutler_kennefick_poisson_1994}, and we will exclude unstable systems from this analysis.

A relativistic anomaly $\chi$, which ranges from 0 to $2\pi$, is defined so that
\begin{equation}
    r(\chi)=\frac{pM_3}{1+e_o \cos\chi},
\end{equation}
Furthermore, the azimuthal coordinate is given by 
\begin{equation}
    \varphi(\chi)=2\Bigr(\frac{p}{p-6+2e_o}\Bigr)^{1/2}\Bigr[F\Bigr(\frac{\chi}{2}+\frac{\pi}{2},k^2\Bigr)-F\Bigr(\frac{\pi}{2},k^2\Bigr)\Bigr]+\gamma_o,
    \label{varphi vs chi}
\end{equation}
where $k^2=\frac{4e_o}{p-6+2e_o}$, $F$ is the incomplete elliptic integral of the first kind, and $\gamma_o$ denotes 
the initial argument of pericenter for the outer orbit (see Note \ref{gamma_o footnote}).

The relationship between time and the relativistic anomaly is given by
\begin{multline}
    t(\chi)= p^2M_3(p-2-2e_o)^{1/2}(p-2+2e_o)^{1/2}\\
    \times\int_0^{\chi}d\chi'\Bigr\{(p-2-2e_o\cos\chi')^{-1}(1+e_o\cos\chi')^{-2}\\
    \times(p-6-2e_o\cos\chi')^{-1/2}\Bigr\}.
\end{multline}

In the end, the geodesic has a doubly periodic structure, and the radius has a period of $r(\chi)$ has a period of $P_r = t(2\pi)$. During a time of $P_r$, however, the azimuthal variable travels further than $2\pi$, which is the relativistic pericenter precession.
It is useful to define the shift in angle over a radial period. This is equal to
\begin{equation}
    \Delta\varphi = 4\Bigr(\frac{p}{p-6+2e_o}\Bigr)^{1/2}F(\pi/2,k^2) .
\end{equation}
We note that this matches the 1PN GR result~\cite{Will:2018mcj} for the amount of precession during a radial period in the limit $p\gg 1$
\begin{equation}\label{eq:1pn-peri-prec}
    \Delta\varphi \approx 2\pi(1+3/p) =6\pi/p + 2\pi\, .
\end{equation}
Defining the  azimuthal frequency $\Omega_\varphi\equiv \Delta\varphi/P_r$, it is shown that $\varphi(t)-\Omega_\varphi t$ is $P_r$--periodic~\cite{cutler_kennefick_poisson_1994}. We note that $\varphi(t)$ itself is \emph{not} periodic -- since the orbit precesses, it takes $<P_r$ time for $\varphi$ to move through $2\pi$ radians. Even though the precession angle over a full orbit remains constant, the time it takes to move through the precession angle will depend on the BBH distance from the SMBH (conserving angular momentum), so for an eccentric orbit, the time to complete a full $2\pi$ in $\varphi$ will depend on the starting value of $\varphi$ itself.

To find $r(t)$ and $\varphi(t)$ numerically over many full orbits, we calculate the orbit over $\chi \in [0,2\pi]$ and utilize the periodicity of $r(\chi)$ and $\varphi(t)-\Omega_\varphi t$.
We furthermore choose some $\chi_0\equiv\chi(t=0)$ so that $\varphi(\chi_0)=\varphi_0$, where $\varphi_0$ is the initial azimuthal coordinate of the BBH in the plane of the outer orbit (see the bottom of Fig. \ref{fig:Triple system geometry}). 
Furthermore, $\dot{r}(t)$ and $\dot{\varphi}(t)$ can be calculated by application of the chain rule to the expressions relating $r$, $\varphi$, and $t$ to $\chi$ above. 

\subsection{Waveform}
We can now proceed to calculate the strain detected by the space-based observatory, using the formalism of \cite{Apostolatos:1994mx}. The overall measured signal is 
\begin{multline}
    \tilde{h}(f)=\tilde{h}_C\sqrt{(A_+F_+)^2+(A_\times F_\times)^2}\\
    \times\exp\{-i[\Phi_P+2\Phi_T+\Phi_D]\}, 
    \label{eq:final strain}
\end{multline}
where $\tilde{h}_C$ is the carrier waveform of the BBH, $A_{+,\times}$ and $F_{+,\times}$ are the polarization amplitude and antenna response, respectively, and $\Phi_P$, $\Phi_D$, and $\Phi_T$ are the polarization, Thomas, and Doppler phases.
The carrier waveform in the frequency domain to leading post-Newtonian (PN) order is \cite{Cutler:1994ys}
\begin{multline}
    \tilde{h}_C(f)=\Bigr(\frac5{96}\Bigr)^{1/2}\frac{\mathcal{M}^{5/6}}{\pi^{2/3}D_L}f^{-7/6}
    \\\times\exp\{i[2\pi ft_c-\phi_c-\frac\pi 4+\frac34(8\pi \mathcal{M}f)^{-5/3}]\},
\end{multline}
where $t_c$ and $\phi_c$ are the time and phase at coalescence.
To the leading PN order, the relationship between GW frequency and time is given by
\begin{equation}
\label{eq:t vs f}
    t(f)\approx t_c-\frac{5}{256\pi^{8/3}}\frac{1}{\mathcal{M}^{5/3}f^{8/3}}.
\end{equation}

The two polarizations of the strain, $h_+$ and $h_\times$, are modified by the amplitude factors
\begin{gather}
\label{eq: aplus}
    A_+=1+(\hat{L}_i\cdot\hat{N})^2\\
\label{eq: across}
    A_\times = -2\hat{L}_i\cdot\hat{N},
\end{gather}
and furthermore, the antenna responses for a 90-degree detector are
\begin{multline}
    \label{eqn:detframe1}
    F_+(\theta_S, \phi_S, \psi_S)=\frac12(1+\cos^2\theta_S)\cos2\phi_S\cos2\psi_S\\
    -\cos\theta_S\sin2\phi_S\sin2\psi_S \, ,
\end{multline}

\begin{multline}
    \label{eqn:detframe2}
    F_\times(\theta_S, \phi_S, \psi_S)=\frac12(1+\cos^2\theta_S)\cos2\phi_S\sin2\psi_S\\
    +\cos\theta_S\sin2\phi_S\cos2\psi_S,
\end{multline}
where 
\begin{equation}
\label{eq:psiS}
    \tan\psi_S(t)=\frac{\hat{L}_i\cdot\hat{z}-(\hat{L}_i\cdot\hat{N})(\hat{z}\cdot\hat{N})}{\hat{N}\cdot(\hat{L}_i\times\hat{z})}.
\end{equation}
Note the use of the detector-frame coordinates in Eqs.~\eqref{eqn:detframe1} and \eqref{eqn:detframe2}. 
For a triangular detector such as LISA or B-DECIGO, the antenna pattern acquires a factor of $\sqrt{3}/2$ and there are two effective detectors \cite{Cutler:1997ta}.

Let us now specify the phases in Eq.~\eqref{eq:final strain}. Since the phases are slowly varying functions of time, the stationary phase approximation is used to convert them into frequency-dependent components via Eq. \eqref{eq:t vs f} -- i.e., for some function $g(t)$ appearing in the time-domain waveform $h(t)$, $g(f)\approx g(t(f))$ \cite{Droz:1999qx}.  The polarization phase is given by
\begin{equation}
    \tan\Phi_P(t)=-\frac{A_\times(t)F_\times(t)}{A_+(t)F_+(t)}.
\end{equation}
The Thomas phase arises from the evolution of the principle $+$--polarization axis \cite{Apostolatos:1994mx}, and thus the inner orbital phase of the two stellar mass BH in the BBH, as the angular momentum $\hat{L}_i$ precesses. It is given by 
\begin{equation}
    \Phi_T(t)=-\int_t^{t_c}dt\Bigr[\frac{\hat{L}_i\cdot\hat{N}}{1-(\hat{L}_i\cdot\hat{N})^2}\Bigr](\hat{L}_i\times\hat{N})\cdot\frac{d\hat{L}_i}{dt}\, .
\end{equation}
The final phase term is the Doppler phase shift, the phase shift induced by the changing distance between the detector and the GW source. There are two contributions to this phase. The first is the contribution from the detector, given at a particular time $t$ by 
\begin{equation}
    \Phi_{D,\text{det}}=2\pi f \times(1\text{ AU})\sin\theta_S\cos(\phi_{det}-\phi_S).
\end{equation}
The other is from the source, which is modulated by the changing orbital radius as well as the inclination of the outer orbit and the position of the BBH in that orbit:
\begin{equation}
    \Phi_{D,\text{src}}=2\pi f\times r\sin\iota_J\sin\varphi.
\end{equation}

Gravitational lensing from the SMBH and its host galactic nucleus is neglected in this waveform, though its effects on parameter estimation have been studied in \cite{Yu:2021dqx,Christian:2018vsi}.

\subsection{Neglected Orbital Dynamics}\label{subsec:neglected orbital dynamics}

A three body system is complicated, and exhibits some interesting phenomenology. We will now discuss several additional well-known behaviors, and why we neglect them. A useful benchmark for comparison is that the characteristic frequency for de Sitter precession scales as
\begin{equation}
    \Omega_\text{dS} = \frac{1}{1.1 \text{ yr}} \bigg(\frac{100}{a_o/M_3} \bigg)^{5/2}\bigg( \frac{10^8 M_\odot}{M_3}\bigg) \bigg( \frac{1-0.3^2}{1-e^2}\bigg) \, .
\end{equation}
We consider the implications of non-zero BH spins on the orbital dynamics. The precession of $\hat L_o$ around the spin of the SMBH $\hat S_3$ with $S_3=\chi_3M_3^2$ has characteristic frequency \cite{Liu:2021uam}
\begin{equation}
    \Omega_{L_o,S_3}=\frac{S_3(4+3(M_1+M_2)/M_3)}{2a_o^3(1-e_o^2)^{3/2}}.
\end{equation}
If we consider the case $M_3 \gg M_1 + M_2$
\begin{equation}
    \frac{1}{t_{L_o,S_3}} = \frac{1}{9.7 \text{ yr}}\bigg( \frac{\chi_3}{0.7}\bigg) \bigg(\frac{100}{a_o/M_3} \bigg)^{3} \bigg( \frac{1-0.3^2}{1-e^2}\bigg)^{3/2}.
\end{equation}
Even for rapidly spinning SMBHs, this effect is about one order of magnitude slower than de Sitter precession, so for now, we neglect it. It is worth noting that each successive effect included in the waveform modulation generally increases the amount of Fisher information. As such, we expect that future inclusion of this effect will lead to further improved parameter estimation uncertainties.

Lense-Thirring precession of $\hat L_i$ around $\hat S_3$ also contributes to the orbital dynamics, with 
\begin{equation}
    \Omega_{L_i,S_3}=\frac{S_3}{2a_o^3(1-e_o^2)^{3/2}}.
\end{equation}
This precession frequency is one-quarter of $\Omega_{L_o,S_3}$, and thus, since we treat $\Omega_{L_o,S_3}$ as small in this work, we do the same for $\Omega_{L_i,S_3}$.

As in \cite{Yu:2020dlm}, we also neglect the precession of $\hat{L}_i$ around the spins of the two stellar mass BH. The opening angle of this precession will be of order $1^\circ$, much less than a typical value of $\lambda_L$ \cite{Tagawa:2020dxe}. Also, the effects of this spin-induced precession should be easily distinguishable from the Doppler shift or de Sitter precession because the spin-induced precession will occur over just days, rather than years, for GW frequencies in the bands of space-based observatories. 

We also consider Lidov-Kozai oscillations, the Newtonian tidal effect which exchanges inner orbit eccentricity with inclination between $\hat L_o$ and $\hat L_i$ \cite{Suzuki:2020zbg}. These oscillations have a characteristic frequency of \cite{Liu:2021uam}
\begin{equation}
    \Omega_\text{LK}=\Omega_i\frac{M_3}{M_1+M_2}\Bigr(\frac{a_i}{a_o\sqrt{1-e_o^2}}\Bigr)^3,
\end{equation}
where $\Omega_i=\sqrt{(M_1+M_2)/a_i^3}$. 
The LK timescale is
\begin{multline}
    \frac{1}{t_\text{LK}} = \frac{1}{67 \text{ yr}}\bigg( \frac{10^8 M_\odot}{M_3}\bigg)^2 \bigg(\frac{100}{a_o/M_3} \bigg)^{3}\\
    \times\bigg( \frac{1-0.3^2}{1-e^2}\bigg)^{3/2} \bigg( \frac{10^{-2} \text{ Hz}}{f}\bigg).
\end{multline}
In our frequency band of interest, this effect occurs over much longer time scales than the de Sitter precession, and since both de Sitter precession and Lidov-Kozai oscillations modulate $L_i$, we neglect the slower of the two processes.

We furthermore assume that the eccentricity of the inner binary $e_i$ is zero. As explained in \cite{Yu:2020dlm}, the inner eccentricity does not affect any component of the measured strain outside of the carrier waveform $\tilde{h}_C(f)$, and thus should influence parameter estimation uncertainties primarily through the SNR. Furthermore, the eccentric Kozai-Lidov mechanism can drive periodic modulation of $e_i$ between moderate and very high values. The GW signal frequency from a stellar-mass BBH can be pushed into the sensitivity range of space-based observatories when the inner eccentricity is high, so the eccentric Kozai-Lidov mechanism can produce periodic high SNR bursts in these detectors, driving up the total SNR measured for that particular binary \cite{knee2023, alaninprep}. However, the time scale of this periodic burst behavior scales roughly as \cite{Naoz2016} $\Omega_o^{-2}f(1-e_o^2)^{3/2}$. These effects therefore occur much more slowly than de Sitter and pericenter precession, and thus are left for implementation into future analyses.

A higher $e_i$ also leads to faster merger times; however, high eccentricity BBHs can still remain in the millihertz and decihertz frequency bands throughout the entire observation period with just a larger initial separation between the two stellar mass BHs. So, it is expected that even for $e_i$ approaching 1, such BBHs will offer long enough integration times to generate a moderate SNR, and therefore the inner eccentricity should not significantly alter the results of the simplified Fisher matrix analysis (see \cite{Yu:2020dlm} for a more detailed discussion).

\section{Parameter Estimation with the Fisher Information Matrix}
\label{sec: simplifications}
In this analysis, we implement the Fisher information matrix method (as done in \cite{Yu:2020dlm}) as a simple estimator for how well properties of a BBH+SMBH triple system can be measured. We make a number of well-supported assumptions to reduce the complexity of the numerical methods used to estimate parameter uncertainties.
\subsection{Parameter Uncertainties from the Fisher Information Matrix}
\label{subsec:FIMatrix}
We first outline how the Fisher information matrix (from now on, Fisher matrix) is used to estimate parameter measurement uncertainties. The elements of the Fisher matrix are defined as
\begin{equation}
    \Gamma_{ab} \equiv \left( \frac{\partial \tilde h(f)}{\partial \theta_a} \Big| \frac{\partial \tilde h(f)}{\partial \theta_b}\right) \, ,
\end{equation}
where 
\begin{equation}
    \left( \tilde g \big|\tilde h \right) = 4 \, \text{Re} \int_0^\infty \frac{\tilde g^{\ast}(f) \tilde h(f)}{S_n(f)} df \, \footnote{We make the approximation that the PSD $S_n(f)$ varies slowly enough so that $S_n(f)$ for the GW frequency in the BBH frame and the Doppler-shifted GW frequency in the observer frame are roughly equal. See App. \ref{sec: doppler shift calculations}.},
\end{equation}
$\tilde h$ is the frequency-domain waveform, $S_n(f)$ is the PSD of the detector noise, and $\theta_a$ are the various parameters of the system. In practice, we limit the frequency bounds of integration to $[f_{min},f_{max}]$, where $f_{max}$ is 
at the upper edge of the detector sensitivity range and $t(f_{max})-t(f_{min})=5$ years (via Eq. \eqref{eq:t vs f}) --  see Sec. \ref{sec: results}.

We note that we use a finite difference method to compute $\partial\tilde{h}/\partial\theta_a$. To choose a finite parameter difference $\Delta\theta_a$ from which to estimate $\partial\tilde{h}/\partial\theta_a$, we minimize the quantity $\epsilon$, analogous to waveform mismatch, 
\begin{equation}
    \epsilon=1-\frac{(\partial_{[\Delta\theta_a]}\tilde{h} | \partial_{[4\Delta\theta_a]}\tilde{h})}{\sqrt{(\partial_{[\Delta\theta_a]}\tilde{h} | \partial_{[\Delta\theta_a]}\tilde{h})(\partial_{[4\Delta\theta_a]}\tilde{h} | \partial_{[4\Delta\theta_a]}\tilde{h})}},
\end{equation}
where 
\begin{equation}
    \partial_{[\Delta\theta]}\tilde{h}=\frac{\tilde{h}(\theta+\Delta\theta)-\tilde{h}(\theta-\Delta\theta)}{2\Delta\theta}.
\end{equation}
Empirically choosing $\Delta\theta_a$ to make $\epsilon$ small gives us the best accuracy in computing the numerical derivative, as $\epsilon$ begins to increase once $\Delta\theta_a$ becomes so small that the changes in $\tilde{h}$ are smaller than computer precision. The choice of $4\Delta\theta_a$ to compare to $\Delta\theta_a$ is arbitrary. 

The Fisher information matrix is related to the covariance matrix roughly by
\begin{equation}
    \Sigma_{ab} = [\Gamma^{-1}]_{ab} + \mathcal{O}(\rho^{-4}),
\end{equation}
where $\rho$ is the signal-to-noise ratio (SNR). So, in the limit of large SNR, the covariance between two parameters $\Delta\theta_i\Delta \theta_j$ is approximately equal to the corresponding element of the inverse of the Fisher information matrix. As such, the parameter estimation uncertainty is given by $\Delta\theta_i=(\Sigma_{ii})^{0.5}$. If a network of GW detectors were to observe the same system, the Fisher information matrix would scale as the sum of the matrix elements for each detector, or
\begin{equation}
    \left( \Gamma_{ab} \right)^\text{network} = \sum_\text{det} \Gamma_{ab}^\text{det}.
\end{equation}
This also applies to a triangular observatory, wherein three arms compose two interferometric detectors.

\subsection{Reduced Fisher Matrix Dimensions}
We can reduce the dimensions of the Fisher matrix by removing certain physical parameters from the analysis. Doing so reduces the total computation time as well as the condition number, leading to improved numerical accuracy in the Fisher matrix inversion \cite{Vallisneri:2007ev}. From the parameters listed in Table \ref{tab:params_table}, our Fisher matrices include the following 12 parameters:
\begin{multline}
    \theta_a = (\log D_L,\overline\theta_S,\overline\phi_S,\overline\theta_J,\overline\phi_J,\lambda_L,\alpha_0,\\\log M_3,\log\Omega_o,\gamma_o,e_o,\varphi_0).
\end{multline}
We can remove parameters which we expect will have strong priors obtained from other GW measurements, or which contribute only weakly to the gravitational waveform. For example, we assume that space-based detectors like LISA or TianGO will act in conjunction with ground-based observatories, which are far more sensitive to the chirp mass $\mathcal{M}$, the mass ratio $q$, and the time and phase of coalescence $t_c$ and $\phi_c$
\cite{Kuns:2019upi}, and thus treat these four parameters as perfectly known in our analysis. Removing the chirp mass from the Fisher matrix also improves the numerical stability of our analysis. Furthermore, we neglect the spins of the three black holes because the precessional effects they induce accumulate much more slowly than the outer orbital motion and de Sitter precession, as described in Sec. \ref{subsec:neglected orbital dynamics}.

\section{Results and Discussion}
\label{sec: results}
\begin{figure}
    \centering
    \includegraphics[width=\linewidth]{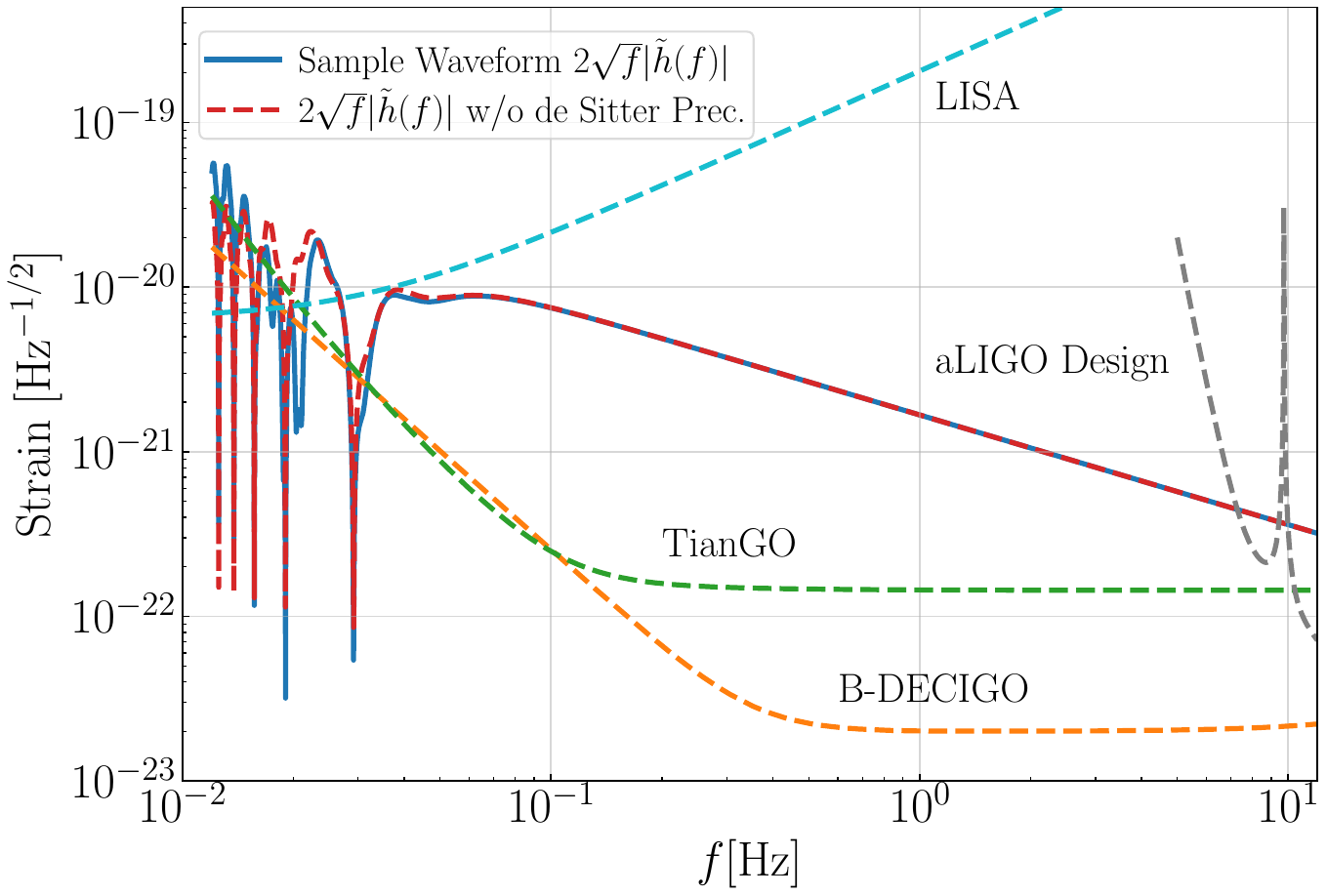}
    \caption{An example waveform $\tilde{h}(f)$ with $M_3=10^8M_\odot$, $a_o=100M_3$, and $e_o=0.3$, along with approximate sensitivity curves for B-DECIGO, TianGO, and LISA used in the Fisher matrix calculations done in this work. The red dashed curve gives the same waveform but with the effects of de Sitter precession removed.}
    \label{fig:waveform_and_sensitivities}
\end{figure}

We examine a BBH+SMBH triple system with fixed parameters $M_1=M_2=50M_\odot$, $t_c=0$, $\phi_c=0$, $D_L=1$Gpc, $(\overline{\theta}_S,\overline{\phi}_S)=(33^\circ, 147^\circ)$, $(\overline{\theta}_J,\overline{\phi}_J)=(75^\circ, 150^\circ)$, and $\lambda_L=45^\circ$. For B-DECIGO, TianGO, and LISA, we compute the Fisher matrix where the integration is taken over a frequency window corresponding to an observation time of five years and the highest frequency is $f_\text{max}=$ 12 Hz -- this roughly corresponds to a lowest frequency of $f_\text{min}\sim 12$ mHz. In Fig.~\ref{fig:waveform_and_sensitivities}, we plot an example frequency-domain waveform along with the B-DECIGO, TianGO, and LISA sensitivity curves used in computing Fisher matrix elements.

\begin{figure}
    \centering
    \includegraphics[width=\linewidth]{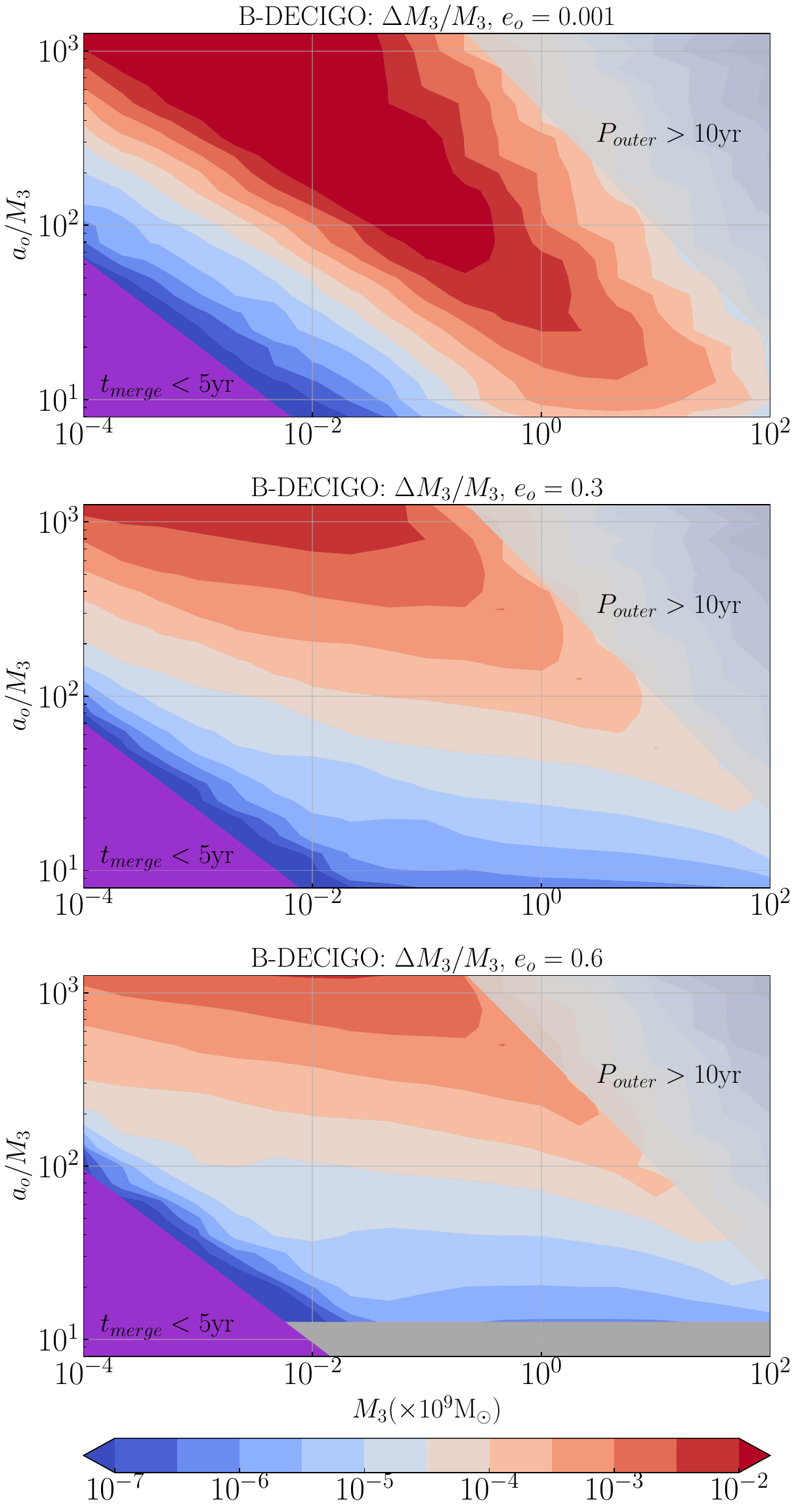}
    \caption{Fractional uncertainty in $M_3$ as measured by B-DECIGO for three different eccentricities $e_o=\{0.001, 0.3, 0.6\}$. At each point in the contour plot, we take the median uncertainty over a set of combinations of $(\gamma_o, \alpha_0, \phi_o)$. The purple region corresponds to where the outer binary merges in less time than the observation duration. We lightly shade out the region with an outer orbital period greater than 10 years, where the cumulative effect of the Doppler shift becomes small.
     }
    \label{fig:BDEC_measure_m3_contours}
\end{figure}

In Fig.~\ref{fig:BDEC_measure_m3_contours}, we plot the fractional uncertainty in the SMBH mass $M_3$, measured by B-DECIGO, as we vary $M_3$ and $a_o$. The Fisher matrix breaks down if $e_o$ is identically zero, so in order to facilitate comparisons to the circular orbits used in \cite{Yu:2020dlm}, we use $e_o=0.001$. At each point, we sample the covariance found with the Fisher matrix over combinations of the three geometrical phases -- that is, 6 choices of $\gamma_o$, $\varphi_o$ and $\alpha_0$, or 216 sets of $(\gamma_o, \alpha_0, \varphi_o)$ -- and find the median. 

The purple regions denote where the outer binary merges in less than the proposed observation length of five years. We expect systems in this region to be exceedingly rare, as there is only a short window for such systems to form in order to be detected by B-DECIGO. We also shade out the region where the outer orbital period $P_\text{outer}$ exceeds twice the observation duration. In this region, the most dominant source of waveform modulation -- namely, the Doppler phase shift -- is difficult to measure because the BBH only passes through a small range of angles over the observation period. Furthermore, when the Doppler phase shift varies slowly, remaining roughly constant over the observation run, it becomes degenerate with $t_c$, which itself can be changed by a simple redefinition of when $t=0$. So, in this shaded region, our assumption that $t_c$ can be safely removed from the list of parameters in the Fisher matrix does not hold well. Indeed, we encounter problems with numerical instability when computing the Fisher matrix in this region of the contour plots.

\begin{figure}
    \centering
    \includegraphics[width=\linewidth]{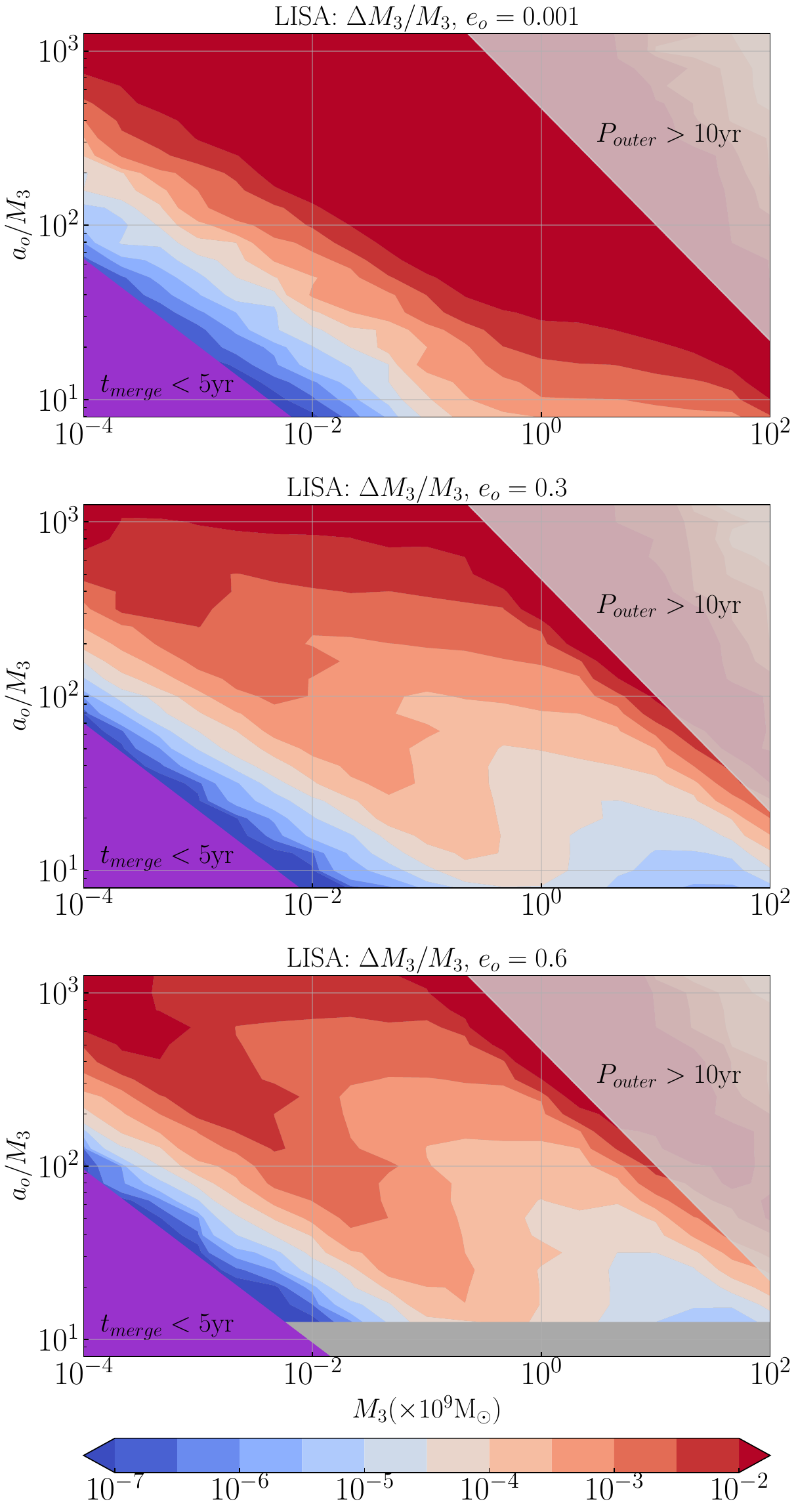}
    \caption{Same as Fig. \ref{fig:BDEC_measure_m3_contours}, but measured by LISA instead.}
    \label{fig:LISA_measure_m3_contours}
\end{figure}

Figure \ref{fig:LISA_measure_m3_contours} gives the same results, but using the LISA detector response and noise curve instead of that of B-DECIGO. The contour plots using the TianGO observatory have a similar structure to those using B-DECIGO, as the two detectors have similar sensitivity curves. Across the majority of the parameter space studied, the two sets of contours differ only in magnitude and not in shape, so for the sake of brevity, they are omitted here. 

We note that the fractional uncertainty in the outer orbit semimajor axis $\Delta a_o/a_o$ follows a similar contour structure to that of $\Delta M_3/M_3$. For the outer orbit, 
\begin{equation}
    3\frac{a_o^3}{M_3}\frac{\Delta a_o}{a_o}\approx \frac{1}{\Omega_o^2}\frac{\Delta M_3}{M_3}-2\frac{1}{\Omega_o^2}\frac{\Delta\Omega_o}{\Omega_o}.
\end{equation} Our calculations determined that across the $(M_3, a_o/M_3)$ parameters space, $\Delta\Omega_o/\Omega_o$ is much smaller in magnitude than $\Delta M_3/M_3$, so
\begin{equation}
    \frac{\Delta a_o}{a_o}\approx\frac{1}{3}\frac{\Delta M_3}{M_3}.
\end{equation}

\begin{figure}
    \centering
    \includegraphics[width=\linewidth]{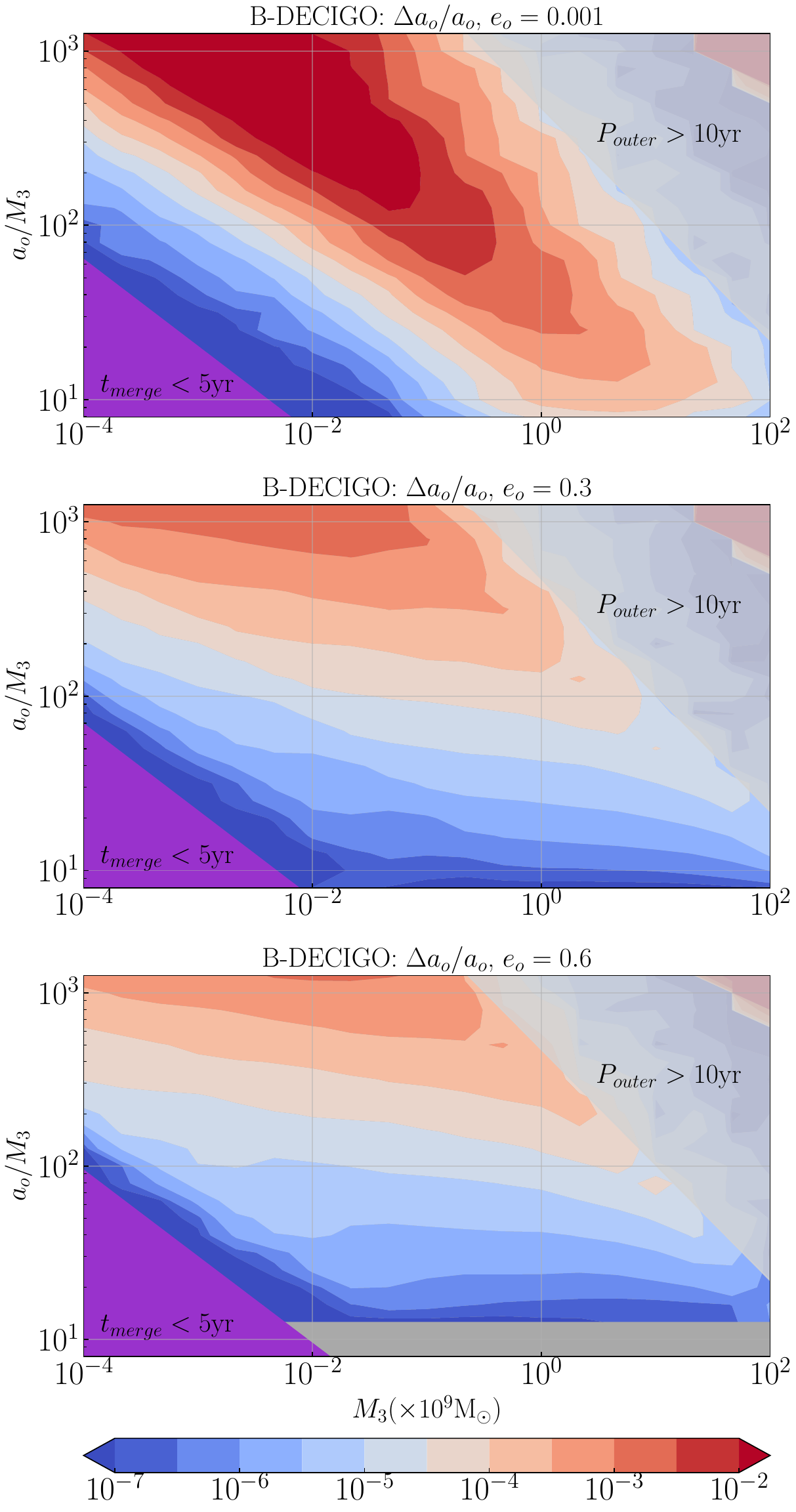}
    \caption{Uncertainty in $a_o$ as measured by B-DECIGO for three different eccentricities $e_o=\{0.001, 0.3, 0.6\}$. The same sampling procedure as used in Fig. \ref{fig:BDEC_measure_m3_contours} is applied here.}
    \label{fig:BDEC_measure_ao_contours}
\end{figure}

\begin{figure}
    \centering
    \includegraphics[width=\linewidth]{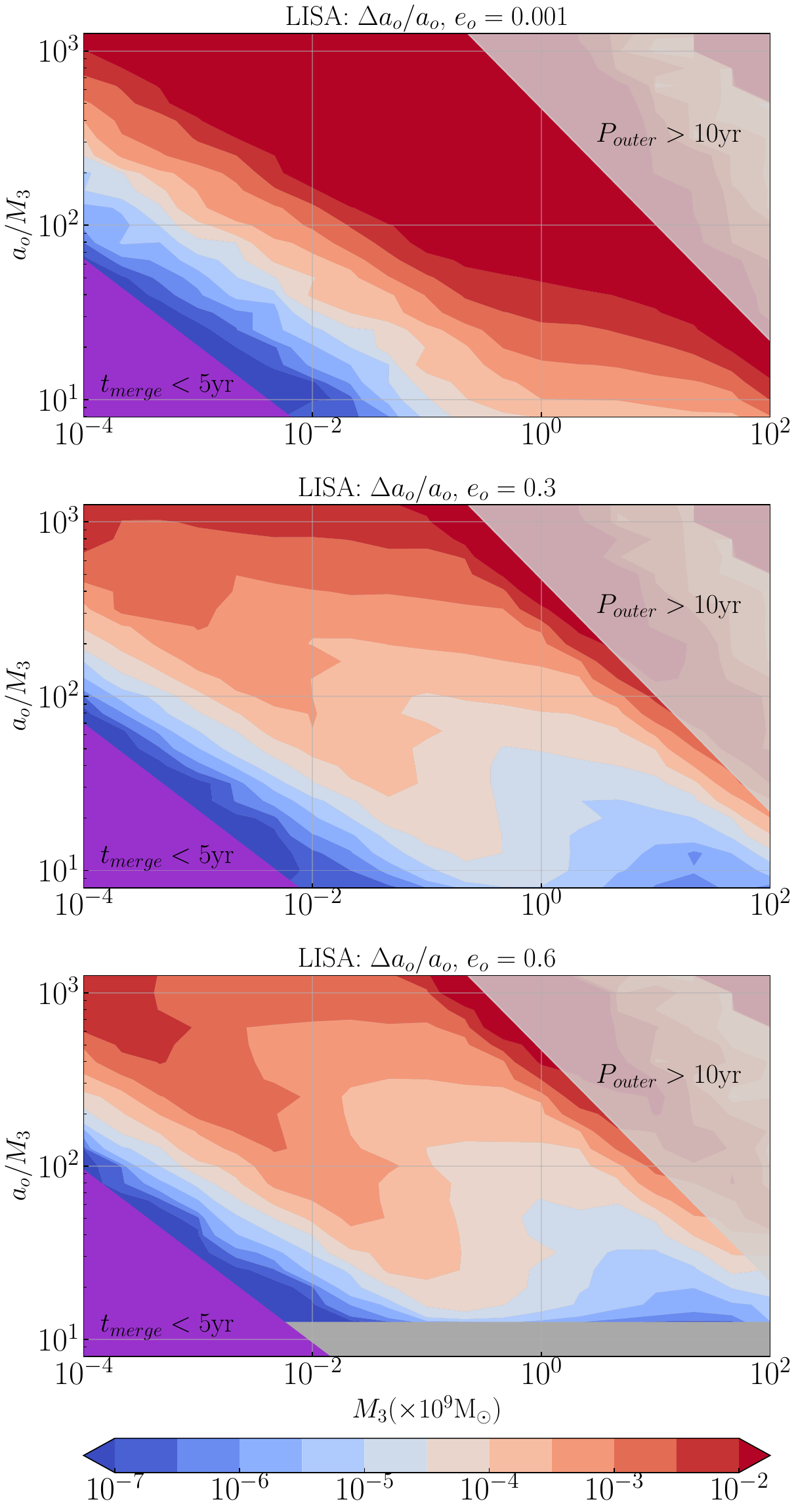}
    \caption{Same as Fig.~\ref{fig:BDEC_measure_ao_contours}, but measured by LISA instead.}
    \label{fig:LISA_measure_ao_contours}
\end{figure}

This result is verified in the structures of Figs. \ref{fig:BDEC_measure_ao_contours} and \ref{fig:LISA_measure_ao_contours}, and we observe that both B-DECIGO and LISA have the potential to realize fractional uncertainties in $M_3$ and $a_o$ significantly below the 0.1\% level across a wide range of parameters of the triple systems.

\begin{figure}
    \centering
    \includegraphics[width=\linewidth]{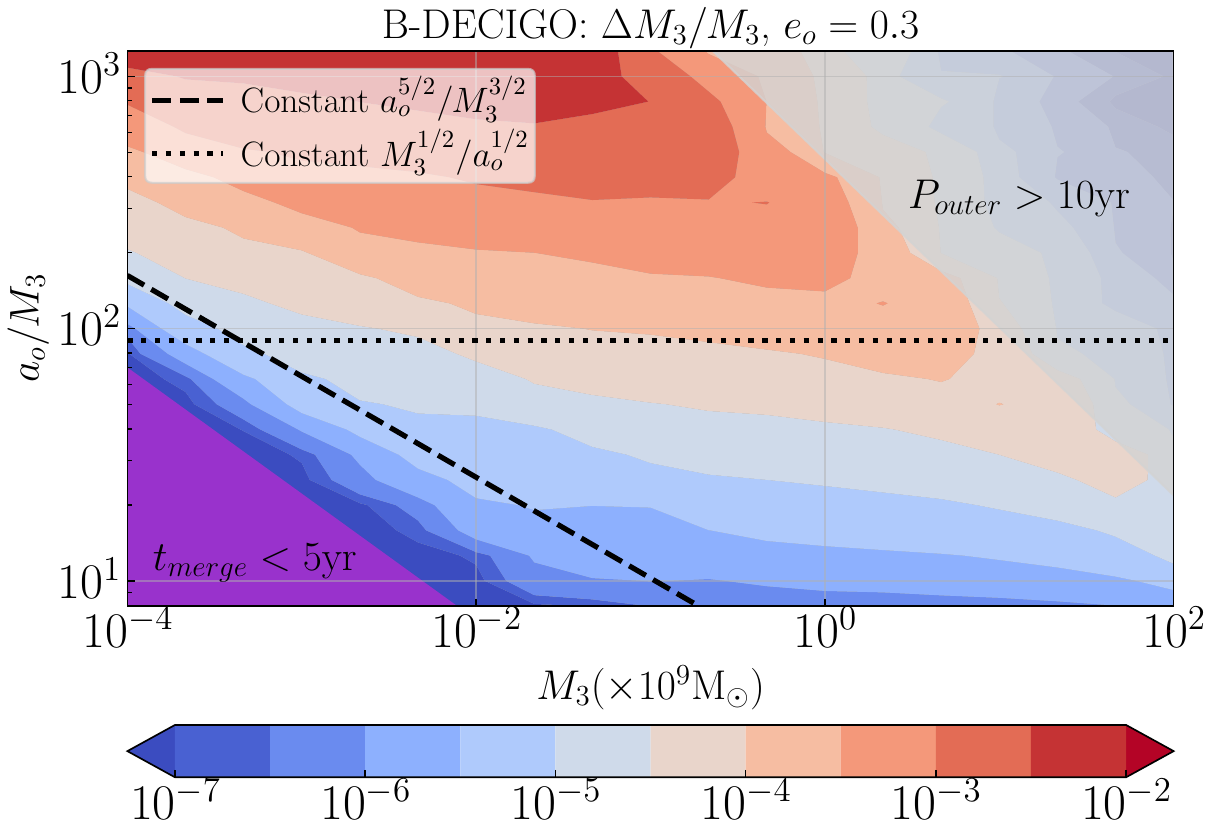}
    \caption{Contour plot for the fractional uncertainty in $M_3$ as measured by B-DECIGO, taken from Fig. \ref{fig:BDEC_measure_m3_contours}. Plotted on top of the contours are lines of constant $a_o^{5/2}/M_3^{3/2}$ and $M_3^{1/2}/a_o^{1/2}$ to indicate the structure of the contours.
    }
    \label{fig:BDEC_m3_ecc_03_labeled}
\end{figure}

To understand the structure of the contour plots, we examine the contour plot in Fig.~\ref{fig:BDEC_m3_ecc_03_labeled}. For small $M_3$ and $a_o/M_3$, the shape of the contours are roughly separated by lines of constant $a_o^5/M_3^3$. We correlate these trends to evolving components of the waveform. 
First, the de Sitter precession frequency is proportional to $\Omega_\text{dS} \propto \sqrt{M_3^3/a_o^5}$. As discussed in App.~\ref{sec:power laws}, the Thomas phase and polarization phase scale as $\Phi_T \sim \Omega_\text{dS} t$. Thus, measurement accuracy scales with the number of de Sitter cycles within the five-year window.
In this region of parameter space, the modulations of de Sitter precession are the dominant effect for how well we can measure $M_3, a_o$.

For larger $M_3$ and a wide range of $a_o/M_3$, the shape of the contours are roughly separated by lines of constant $a_o/M_3$. In this region, the Doppler phase is the dominant term in the frequency domain waveform phase. The Doppler phase magnitude features a degeneracy between $a_o$ and $\sin\iota_J$ (with $\sin\iota_J$ being a function of the angles $\overline\theta_S$, $\overline\phi_S$, $\overline\theta_J$, and $\overline\phi_J$), as these quantities appear in the magnitude only as the product $a_o\sin\iota_J=M_3^{1/3}\Omega_o^{-2/3}\sin\iota_J$. This degeneracy is broken by the inclusion of relativistic pericenter precession, as this produces different periods in the radial and azimuthal motion of the BBH in the outer orbit (cf. Sec. \ref{sec:bbh orbit}). The inclusion of this precession produces lines of constant $\Delta M_3/M_3$ that scale roughly with $(a_o/M_3)^{3/2}$. See App. \ref{sec:power laws} for more detailed discussion.

Studying Fig.~\ref{fig:BDEC_measure_m3_contours}, we see that for $e_o\approx 0$, these flat contours do not appear, as for a circular orbit, pericenter precession is essentially consistent with an increase in $\Omega_o$. The resulting contour plot shape is similar to the results seen in Fig.~5 of \cite{Yu:2020dlm}, where $e_o$ is assumed to be zero -- over a wide range of the parameter space, de Sitter precession is the dominant effect in determining $\Delta M_3/M_3$. However, once $e_o>0$, pericenter precession, rather than de Sitter precession, becomes the leading contribution to $\Delta M_3/M_3$ over a significant portion of the parameter space. The importance of pericenter precession is further emphasized by comparing the magnitudes of $\Delta M_3/M_3$ in our plots to Fig. 5 of \cite{Yu:2020dlm}, which sets $e_o=0$ and therefore does not include pericenter precession (though it does include all other effects used in this work). With pericenter precession included, the parameter uncertainties across a wide region of the overall parameter space can drop by multiple orders of magnitude. 

\begin{figure}
    \centering
    \includegraphics[width=\linewidth]{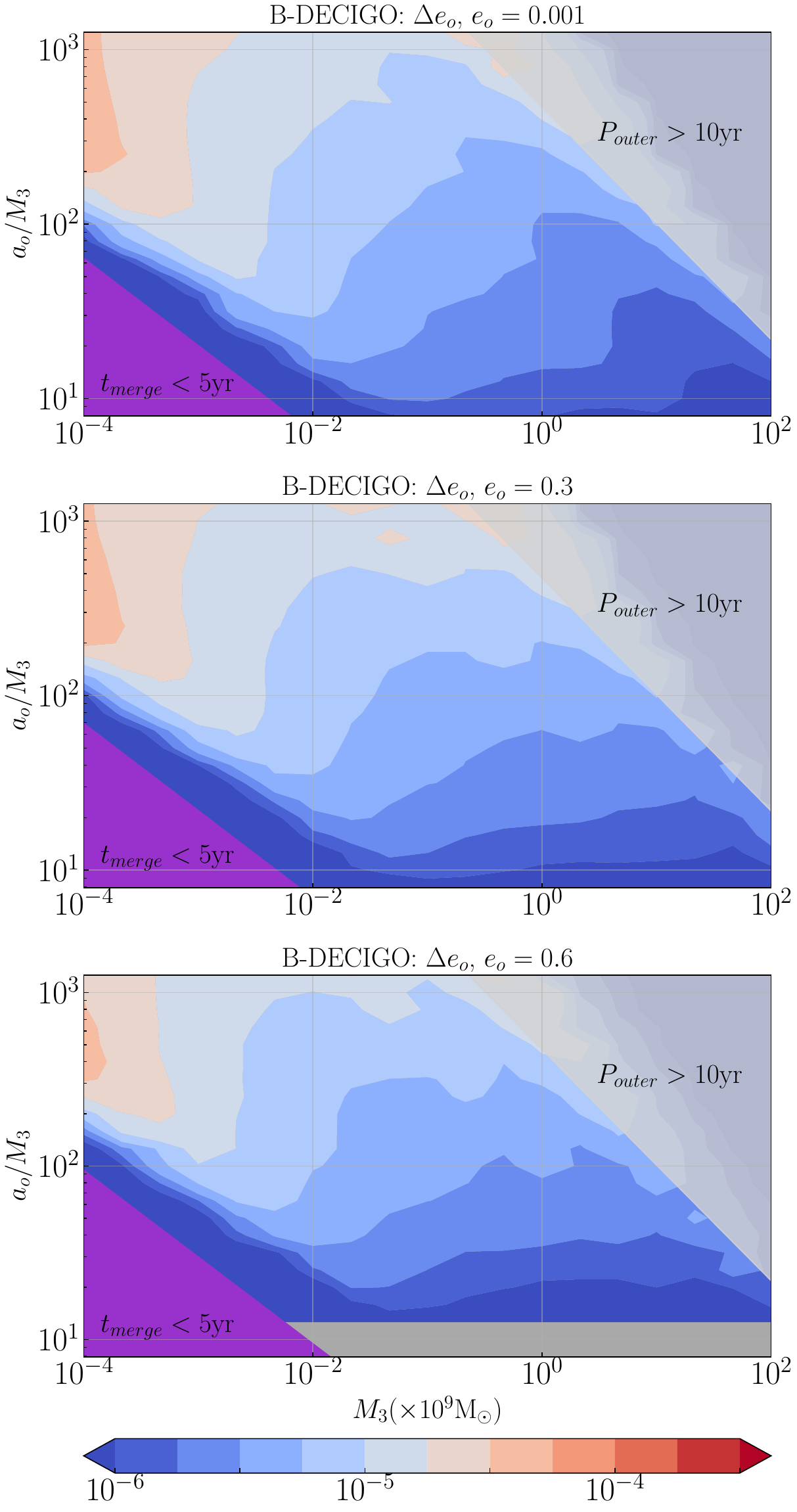}
    \caption{Uncertainty in $e_o$ as measured by B-DECIGO for three different eccentricities $e_o=\{0.001, 0.3, 0.6\}$. The same sampling procedure as used in Fig. \ref{fig:BDEC_measure_m3_contours} is applied here.}
    \label{fig:BDEC_measure_eo_contours}
\end{figure}

\begin{figure}
    \centering
    \includegraphics[width=\linewidth]{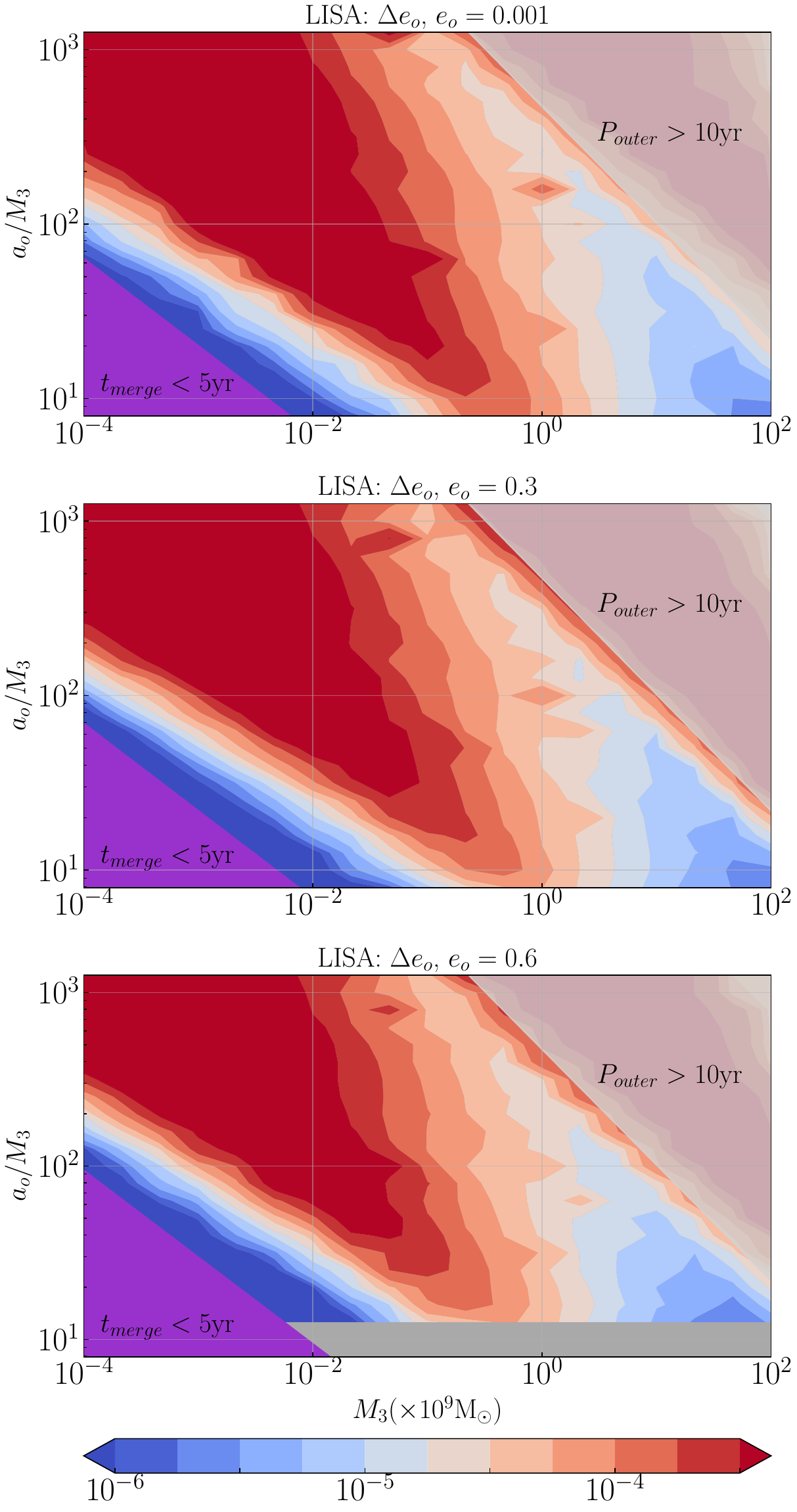}
    \caption{Same as Fig.~\ref{fig:BDEC_measure_eo_contours}, but measured by LISA instead.}
    \label{fig:LISA_measure_eo_contours}
\end{figure}

We also estimate how well the eccentricity can be measured with B-DECIGO and LISA as shown in Fig.~\ref{fig:BDEC_measure_eo_contours} and Fig.~\ref{fig:LISA_measure_eo_contours}.
These results suggest that the eccentricity can be constrained to high precision, with B-DECIGO able to achieve a lower bound of $\Delta e_o\sim 10^{-6}-10^{-5}$ and LISA able to achieve $\Delta e_o\sim 10^{-5}-10^{-4}$ across a substantial portion of the parameter space where precession is detectable. 
Once again, we see the importance of de Sitter precession in the measurability of this parameter -- in the portion of the parameter space where de Sitter precession is rapid, equivalent estimation uncertainties match contours of equal de Sitter precession period. Unlike the contour plots for $\Delta M_3/M_3$, the shape of these contours is not heavily dictated by power laws related to pericenter precession. Indeed, there are no degeneracies between $e_o$ and other waveform parameters which are broken by pericenter precession.

\begin{figure}
    \centering
    \includegraphics[width=\linewidth]{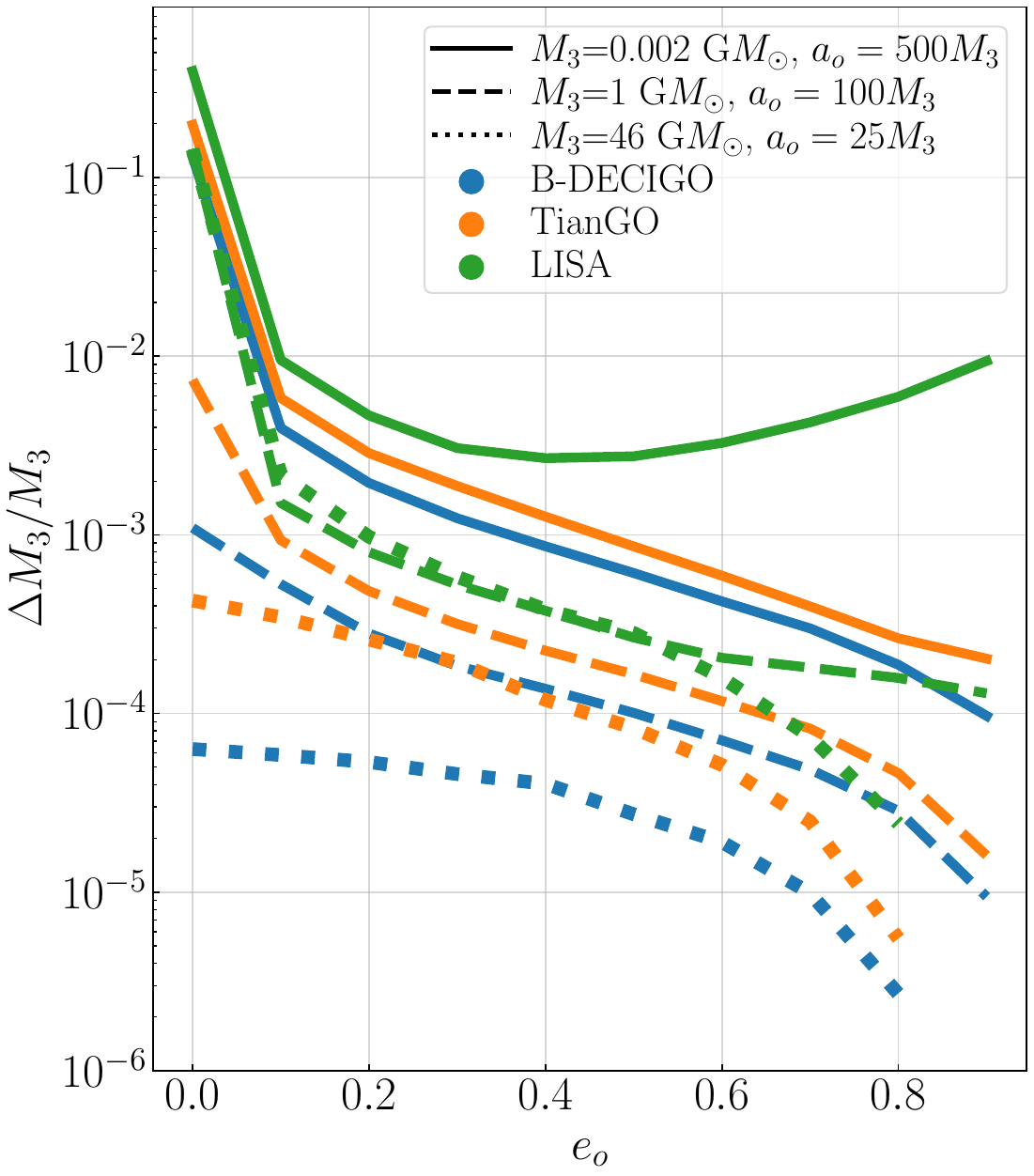}
    \caption{The fractional uncertainties in $M_3$ obtainable by B-DECIGO (blue), TianGO (orange), and LISA (green) as the eccentricity is varied. The solid, dashed, and dotted lines correspond to different choices of $(M_3,a_o/M_3)$.
    }
    \label{fig:Detectors vs m3}
\end{figure}

\begin{figure}
    \centering
    \includegraphics[width=\linewidth]{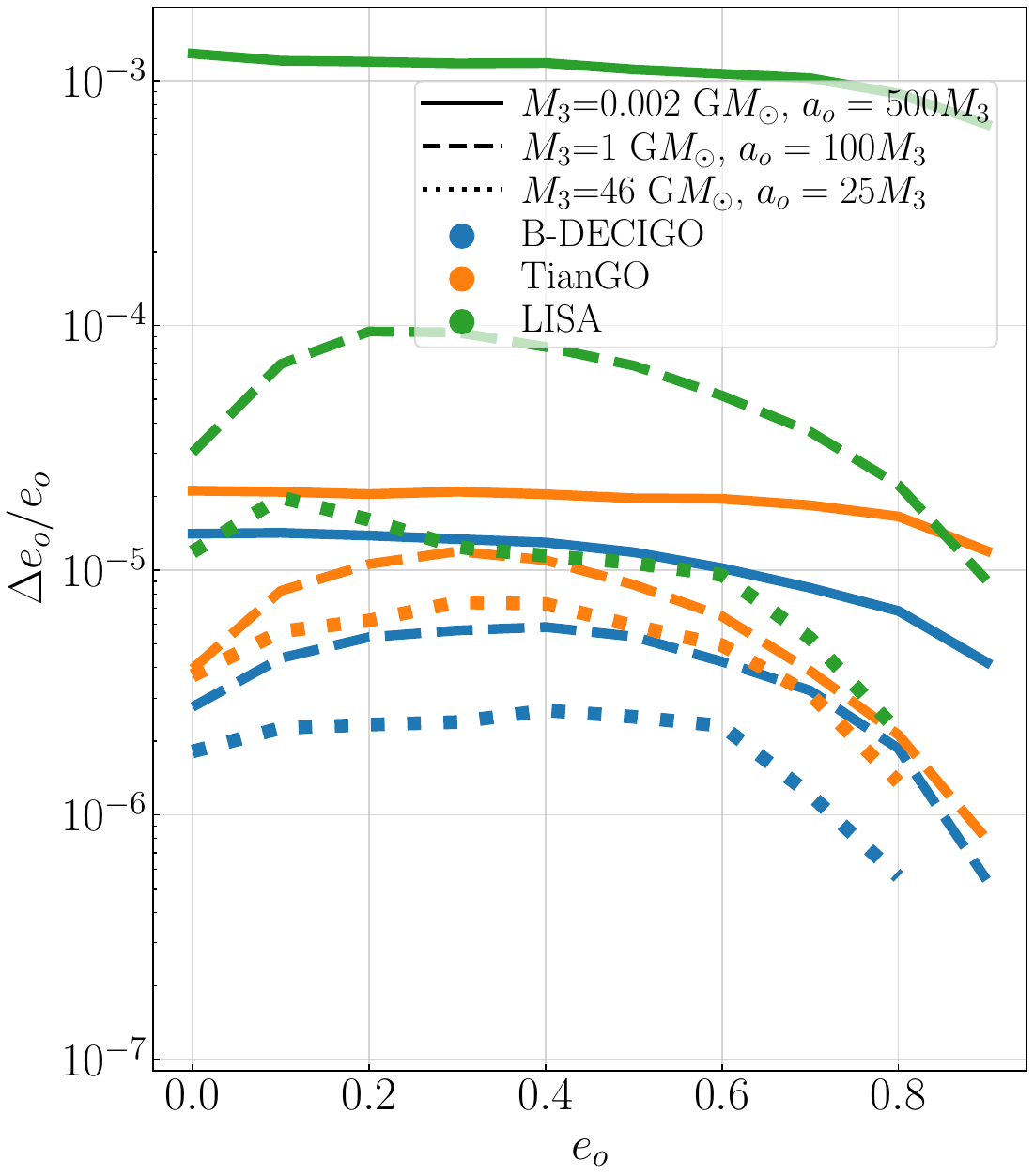}
    \caption{Same as Fig. \ref{fig:Detectors vs m3}, but estimating the outer eccentricity variance $\Delta e_o$.}
    \label{fig:Detectors vs eo}
\end{figure}

An important question is the impact of increasing outer orbit eccentricity on the ability to measure parameters like $M_3$, $a_o$, and $e_o$ itself. In Figs.~\ref{fig:Detectors vs m3} and \ref{fig:Detectors vs eo}, we consider B-DECIGO, LISA, and the TianGO concept and three different combinations of $(M_3, a_o/M_3)$ across our chosen parameter space.
We study the effect of increasing eccentricity on the estimation uncertainties in $M_3$ and $e_o$ (still averaging over initial orbital angles) and find that increasing eccentricity can produce marginal improvements in the measurement of $M_3$ and $e_o$ -- a factor of $\sim$ a few -- though such improvement is not universal across $(M_3, a_o/M_3)$ parameter space.

Considering the arguments given in App.~\ref{sec:power laws}, we see that the leading contributions to the Fisher matrix elements come from the derivatives of $\Phi_D$, $\Phi_P$, and $\Phi_T$. Noting that these phases evolve at secular rates of $\Omega_\text{dS}$ (for $\Phi_P$ and $\Phi_T$) or $\Omega_\text{pericenter}=\Omega_o\frac3p$ (for $\Phi_D$ -- specifically, this is the rate at which the degeneracy between $a_o$ and $\sin\iota_J$ is broken), and recalling that these rates scale with $(1-e_o^2)^{-1}$, it follows that larger eccentricities produce more rapid evolution, larger Fisher matrix entries, and ultimately smaller parameter uncertainties.

The relative sensitivities between the three detectors are responsible for the clear hierarchy in the parameter uncertainties they produce. For example, the rates of precession and orbital velocity are sensitive to both $M_3$ and $e_o$ but with different dependencies, so there exist degeneracies between these two parameters. These degeneracies can be lifted by observing the system over long periods of time so that these effects can accumulate, enabling tighter constraints on their respective individual rates. Examining Fig.~\ref{fig:waveform_and_sensitivities}, we see that LISA effectively measures the BBH signal over a smaller frequency band than the other two detectors in the five years prior to merger. Since the LISA sensitivity is poorer than the other two detectors in the frequencies sampled in the five year observation run, the SNR of the waveform is reduced and it becomes more difficult to extract the waveform modulations driven by orbital and precessional effects over that period of time. Therefore, the degeneracies are not as cleanly lifted in LISA measurements, especially when these rates are slow (i.e., low $M_3$, high $a_o/M_3$), producing less precise parameter estimates. 

The primary effect of eccentricity then is to increase the strength of waveform modulations by increasing the magnitude of the precessional effects (pericenter, de Sitter); however, we see that for the LISA observatory, the improvement in parameter estimation uncertainty with rising $e_o$ is not as significant as in B-DECIGO and TianGO, and in some cases, a larger $e_o$ produces larger uncertainties. While increasing the eccentricity boosts the orbit averaged rate of de Sitter and pericenter precession (Cf. Eqs. \ref{eq:Omega_dS Orbit Averaged} and \ref{eq:1pn-peri-prec}), the majority of this evolution occurs when the BBH is near the outer orbit pericenter and the instantaneous precession rate is largest. So, for systems with slow outer orbits (once again, low $M_3$ and high $a_o/M_3$), an increasing eccentricity constrains the majority of the waveform modulation effects to a shorter time window, as the BBH passes through the region near the pericenter at a faster rate. The GW radiation from the BBH then evolves through a smaller range of frequencies while the waveform is significantly modulated. 

\section{Conclusion and Future Directions}
\label{sec: conclusions}
Using the Fisher information matrix, we have shown that future space-based GW observatories may be able to precisely constrain the properties of BBH+SMBH triple systems, like the SMBH mass and outer orbit semimajor axis and eccentricity, through the GW signal observed from the BBH. We have demonstrated that the rate of change of the Doppler phase shift and the de Sitter precession rate are the dominant factors determining of the measurability of triple system parameters and that an increasing outer orbit eccentricity leads to improved measurement uncertainties through greater Doppler phase shift modulation and faster de Sitter precession. We have also shown that the planned LISA detector is capable of measuring these systems, though decihertz detector concepts such as TianGO or B-DECIGO would possess a competitive advantage over LISA in measuring such quantities.

There are some important limitations of the Fisher information method implemented in this work. As described in \cite{Vallisneri:2007ev}, a high SNR is required for the inverse Fisher matrix to give the covariance of the posterior probability distribution for the true source parameters $\vec\theta_0$. While the SNR we compute for our waveform is generally $\sim40$ for TianGO, it is only $\sim 4$ for LISA, suggesting that the true parameter estimation uncertainties may be significantly different than those calculated here. However, the inverse Fisher matrix is also a \emph{lower bound} for the uncertainty of an unbiased estimator of $\vec\theta_0$ \cite{Vallisneri:2007ev}, so our results essentially offer a best-case scenario for the parameter estimation precision obtainable by future space-based observatories. A more thorough approach to this analysis will implement a full Bayesian methodology.

We can further develop this work by inclusion of additional effects into the waveform. One can implement the spin-precession effects that we chose to neglect in \ref{subsec:neglected orbital dynamics} due to their significantly slower time scales. Furthermore, for triple systems with lower outer binary merger times (i.e., with $M_3$ and $a_o/M_3$ near the purple regions shown in the contour plots such as Fig. \ref{fig:BDEC_measure_m3_contours}), the semimajor axis and outer eccentricity can evolve significantly in time due to radiation reaction \cite{Maggiore:1900zz}. Considering the frequency integral that composes the Fisher matrix elements, we can include the effects of gravitational redshift and Doppler frequency shift, which would require the waveform and detector sensitivity to be evaluated at different frequencies in the integrand. Also, the stationary phase approximation used in the frequency domain waveform  (outlined in App. \ref{sec: doppler shift calculations}) may not hold well for highly eccentric outer orbits, as the outer orbital angle varies quite rapidly near the pericenter for such orbits.

In Ref. \cite{Yu:2021dqx}, it is discussed how gravitational lensing of GWs by the SMBH combined with the de Sitter precession of $\boldsymbol{L}_{\mathrm{i}}$ can further constrain the parameters of a triple system as estimated by a space-based GW observatory, even in the case of a circular outer orbit. It would be interesting to examine the combined effects of an eccentric outer orbit and repeated GW lensing in parameter estimation problems.

Finally, measurements of the motion of a BBH through space through its modulated waveform may prove useful for understanding phenomena besides BBH+SMBH hierarchical triples. For example, measuring the evolving Doppler shift and aberrations induced by the evolving position and velocity of an isolated BBH might enable estimates of BBH kicks that occur shortly before merger or improve the precision of estimates of the Hubble constant by further constraining the redshifts of GW standard sirens \cite{Torres-Orjuela:2020cly, Torres-Orjuela:2022uuy}.

\begin{acknowledgements}
This work was supported in part by the Caltech LIGO Summer Undergraduate Research Fellowship and by the REU program of the NSF.
B.S. acknowledges support by the National Science Foundation Graduate Research Fellowship under Grant No. DGE-1745301. H.Y. is supported by NSF PHY-2308415. 
\end{acknowledgements}

\bibliography{refs}
\newpage

\appendix
\section{Validity of Waveform Approximations}
\label{sec: doppler shift calculations}
We consider some approximations that are made in the formulation of the frequency domain waveform. 
For the source frame waveform $h_s(t_s)=A_s(t_s)e^{-i\Phi_s(t_s)}$, we define the connection between time and gravitational wave frequency in the source frame by
\begin{equation}
    \dot\Phi_s(t_s)=2\pi f_s.
\end{equation}
The outer orbit of the BBH induces a change in the time at which a GW of a particular frequency reaches a fixed observer, which we denote $t_o$. Clearly marking the dependencies of various times on one another,
\begin{equation}
    t_o(t_s)=t_s+t_\parallel(t_s)+D_L/c,
\end{equation}
where $t_\parallel=a_o\sin\iota_J\sin\varphi$ is the time is takes for a radiated GW to propagate from the BBH to the SMBH along the direction of $\hat N$, assuming a circular outer orbit.

With $h_s$ and $h_o$ being the strain in the source and observer frames, we then have
\begin{equation}
    h_o(t_o)=h_s(t_s)=h_s(t_o-t_\parallel-D_L/c),
\end{equation}
which in the Fourier transform (as a function of the observed frequency $f_o$) becomes
\begin{equation}
    \tilde{h}_o(f_o)=\int h_s(t_o-t_\parallel-D_L/c)e^{2\pi if_ot_o}dt_o.
\end{equation}
Inputting the form of $h_s(t_s)$ gives
\begin{equation}
    \tilde{h}_o(f_o)=\int A_s(t_o-t_\parallel-D_L/c)e^{-i\Phi_s(t_o-t_\parallel-D_L/c)}e^{2\pi if_ot_o}dt_o,
\end{equation}
and assigning $t_o-D_L/c=t$ so that $t_s=t-t_\parallel$ produces
\begin{equation}
\label{eq: h_o t-t_parallel}
    \tilde{h}_o(f_o)=e^{2\pi if_oD_L/c}\int A_s(t-t_\parallel)e^{-i\Phi_s(t-t_\parallel)}e^{2\pi if_ot}dt.
\end{equation}

For a typical system we study (e.g., $M_3=10^8M_\odot$, $a_o=100M_3$), $t_\parallel\sim10^4-10^5$ seconds (depending on the orbital angle) and $\dot{t}_\parallel\sim t_\parallel\Omega_o\sim0.01-0.1$.

We can make a number of simplifications to this expression. First, $\Phi_s(t-t_\parallel)\approx \Phi_s(t)-2\pi f_st_\parallel$ as long as $\dot{f}_st_\parallel \ll f_s$. For an inspiral regime BBH with two 50$M_\odot$ BH, $f_s/\dot{f}_s\sim 3\times 10^{3}f_s^{-8/3}$s \cite{Maggiore:1900zz}, and with the majority of the time-integration taking place with $f_s \lesssim 0.1$Hz, the approximation using $\dot{f}_st_\parallel \ll f_s$ holds well.

We apply a similar approximation to $A_s(t-t_\parallel)$. The time scale for the evolution of the GW amplitude is roughly \cite{Cutler:1994ys,Maggiore:1900zz} $A_s/\dot{A}_s\sim\frac32\frac{f_s}{\dot f_s}\gg t_\parallel$, so we can reasonably approximate $A_s(t-t_\parallel)\approx A_s(t)$. This simplifies Eq. \eqref{eq: h_o t-t_parallel} to
\begin{equation}
    \tilde{h}_o(f_o)\approx e^{2\pi if_oD_L/c}\int A_s(t)e^{-i[\Phi_s(t)-2\pi f_st_\parallel(t)]}e^{2\pi if_ot}dt.
\end{equation}
Consider an expansion of $F(t)=\Phi_s(t)-2\pi f_st_\parallel(t)$ around some time $t'$. Noting that $t_\parallel$ is a function of $t$, we find
\begin{multline}
    \Phi_s(t)-2\pi f_st_\parallel(t)\approx\Phi_s(t')-2\pi f_st_\parallel(t')\\+2\pi(f_s-f_s\dot{t}_\parallel-\dot{f}_st_\parallel)(t-t')+\frac12\ddot{F}(t')(t-t')^2+...
\end{multline}
Since $\dot{f}_s/f_s\ll \dot{t}_\parallel/t_\parallel$, the linear term approximates to $2\pi f_s(1-\dot{t}_\parallel)$. Now, to complete the Fourier transform, we turn to the stationary phase approximation. Namely, the majority of the integral comes from the region where the argument of the oscillating term is stationary, which occurs at a time $\tau$ when $f_o-f_s(\tau)(1-\dot{t}_\parallel(\tau))=0$.

Inserting the second order expansion found above into the Fourier transform gives
\begin{multline}
    \tilde{h}_o(f_o)\approx e^{2\pi if_oD_L/c}\int A_s(\tau)
    \\\exp\Bigr\{-i[-2\pi f_ot+\Phi_s(\tau)-2\pi f_s(\tau)t_\parallel(\tau)\\+2\pi f_s(\tau)(1-\dot{t}_\parallel(\tau))(t-\tau)\\+\frac12\ddot{F}(\tau)(t-\tau)^2+...]\Bigr\}dt,
\end{multline}
which simplifies to
\begin{multline}
\label{eq:simplified fourier}
    \tilde{h}_o(f_o)\approx e^{2\pi i[f_oD_L/c+f_s(\tau)t_\parallel(\tau)]}e^{2\pi if_s(\tau)(1-\dot{t}_\parallel(\tau))\tau}\\
    \int A_s(\tau)e^{-i[\Phi_s(\tau)+\frac12\ddot{F}(\tau)(t-\tau)^2+...]}dt,
\end{multline}
recalling that $f_s(\tau)(1-\dot{t}_\parallel(\tau))=f_o$.

The expression $\ddot{F}$ has terms proportional to $\dot{f}_s$, $\ddot{f}_st_\parallel$, $\dot{f}_s\dot{t}_\parallel$, and $f_s\ddot{t}_\parallel$. For our typical system, the three latter terms are generally much smaller than the first, allowing a reasonable approximation of $\ddot{F}\sim\ddot{\Phi}_s$.

Carrying out this integral is a standard exercise as in \cite{Bender1999}, and we see that Eq. \eqref{eq:simplified fourier} evaluates to
\begin{multline}
\label{eq:final waveform approx}
    \tilde{h}_o(f_o)\approx \frac{\sqrt{\pi}}{2}e^{2\pi if_oD_L/c+i\dot{F}(\tau(f))\tau-iF(\tau(f))+i\pi/4}\\
    \times A_s(\tau(f))\Bigr(\frac{2}{\ddot{F}_s(\tau(f))}\Bigr)^{1/2},
\end{multline}
recalling $F=\Phi_s-2\pi f_st_\parallel$ and noting that Eq. \eqref{eq:t vs f} can be used to convert time-dependent quantities into frequency dependent ones. It is clear that the Doppler shift is built directly into this definition -- the components of the source radiation (e.g. $A_s$, $\ddot{F}\approx\ddot{\Phi}_s$) which appear in the observed strain are evaluated at a time $\tau$ when the source radiation is emitted at frequency $f_s=f/(1-\dot{t}_\parallel)$. The ratio of the observed frequency to source frequency matches the expansion of the exact form of the Doppler shift given in Eq. \eqref{eq:doppler} below (in the low $\dot{t}_\parallel$ limit). 

This result poses a problem for carrying out Fisher matrix calculations. The Fisher matrix formalism relies on integration over the observed frequencies; while there exists a monotonic relation between the source GW frequency and time, the inclusion of the Doppler frequency shift results in the same observed frequency originating from multiple distinct source frequencies. Furthermore, the PSD term must be evaluated at the observed frequency, while standard results for the GW signal in the frequency domain are parameterized by the source frequency. Without a one-to-one relationship between the observed frequency and source frequency, carrying out the Fisher matrix calculations requires careful attention to these subtleties when evaluating the frequency domain integrand. Future iterations of this analysis will more carefully implement the Doppler shift in comparing the GW signal in the source and observer frames.

The best path forward we can take to produce a reasonable approach for the Fisher matrix computations is to make the approximation that $\dot{t}_\parallel\ll1$. With this simplification (which holds fairly well across the majority of our parameter space, as we describe below), $f_{obs}\approx f_{src}$ and thus the following simple relation emerges:
\begin{equation}
    \tilde{h}_o(f)\approx e^{2\pi if(D_L/c+t_\parallel)}\tilde{h}_s(f).
\end{equation}

Let us more completely explain why we can reasonably neglect the change in GW frequency due to the Doppler shift induced by the BBH orbital velocity around the SMBH. The (exact) longitudinal Doppler shifted frequency is given by
\begin{equation}
\label{eq:doppler}
    f_{obs}=f_{src}\frac{\sqrt{1+\beta_\parallel}}{\sqrt{1-\beta_\parallel}},
\end{equation}
where $f_{obs}$ and $f_{src}$ are the observed and source frequencies, and $\beta_\parallel=v_\parallel/c=\dot{t}_\parallel$ is the source velocity along the line of sight. Using the methods of Sec.~\ref{sec:bbh orbit}, the maximum orbital velocity occurs at pericenter, with
\begin{equation}
    \beta=\frac{p-2-2e_o}{p\sqrt{(p-2)^2-4e_o^2}}(1+e_o)\sqrt{p-6+2e_o}.
\end{equation}
Depending on the argument of pericenter, the magnitude of the source's line of sight velocity 
can reach up to this value. In the parameter space we study, this velocity is maximized over eccentricity when $e_o=0.9$. At this eccentricity, the maximum velocity over semimajor axes in our parameter space occurs when $p\approx 12$ with $\beta_{max}\approx 0.37$. Then, the largest increase in GW frequency due to the Doppler shift is roughly 50\%. 

Making the approximation that $f_{obs}\approx f_{src}$ can produce some inaccuracies in the distance-accumulated phase terms in Eq. \eqref{eq:final waveform approx}; however, because the Fisher matrix formalism includes integration of the product of the waveform and its complex conjugate, the value of the accumulated phase does not have any effect on the results. Furthermore, the sensitivity curves in Fig. \ref{fig:waveform_and_sensitivities} vary slowly in frequency, so therefore, our approximation that the source frame and observer frame GW frequencies are roughly equal does not significantly affect our Fisher matrix calculations and resulting parameter uncertainties.

Future work may implement more rigorous treatment of the waveform in the frequency domain, including corrections suggested in \cite{Chamberlain:2018snj}, for example. However, we expect that the information provided by examining the shifts in observed GW frequencies, which track the BBH orbital velocity, is essentially already provided by the Doppler phase, which tracks the BBH orbital position. As such, we anticipate little improvement in parameter measurability by including this additional effect. 

\section{Description of Measurement Accuracy}
\label{sec:power laws}

In Fig.~\ref{fig:BDEC_m3_ecc_03_labeled}, we note two power laws for the measurement accuracy of $M_3$. We say that the pericenter precession gives $(a_o/M_3)^{3/2}$, while de Sitter precession scales like $a_o^{5/2} /M_3^{3/2}$. Below, we give scaling arguments to understand this plot. As described in Sec. \ref{sec: results}, we measure $a_o \times \sin\iota_J$ and $\Omega_o$ very well via the Doppler shift
\begin{equation}
    \Phi_D = 2 \pi f \hat N \cdot \vec r = 2 \pi f r \sin \iota_J \sin \phi \, ,
\end{equation}
however we need an additional effect to break the degeneracy between $a_o$ and $\sin\iota_J$. In this appendix, we will show how the effect of precession can be understood as separating the radial and azimuthal periods $P_r \approx P_\phi -\# \frac{M}{a}$, and breaks the degeneracy. We will also explain how the Thomas phase and polarization phase terms allow us to also break the degeneracy.

Let us study a simple analytic toy model for precession where the radial period is shortened by a 1PN term. We will consider only a waveform with Doppler phase here to extract the physical reason that pericenter precession helps us measure the SMBH's orbit. We set
\begin{equation}
    \tilde{h}(f)=\exp\Bigr[i(\Phi_D+2\pi ft_c)\Bigr],
\end{equation}
and provide the following simple dynamics to the outer orbit:
\begin{equation}
    \Phi_D(t)=2\pi f\sin\iota_J\frac{a_o(1-e_o)^2}{1+e_o\cos\xi_r}\sin(\xi_\varphi),
\end{equation}
\begin{equation}
    \xi_r=\xi_{r,0}+\Omega_o\left(1-\epsilon\frac{3M_3}{a_o(1-e_o^2)}\right)t,
\end{equation}
\begin{equation}
    \xi_\varphi=\xi_{r,0}+\xi_{\varphi,0}+\Omega_o t\, ,
\end{equation}
where the variable $\epsilon$ is a counting parameter for precession, and set to 0 or 1 at the end of the calculation.
This system gives a rough approximation for a low eccentricity orbit that includes relativistic pericenter precession, assuming $1/\Omega_o\ll t_{obs}$ so that many orbits are completed during $t_{obs}$ and thus the angular velocities for the orbital and precessional motion average out to their secular values. The quantities $\xi_{r,0}$ and $\xi_{\varphi,0}$ are analogous to $\phi_0$ and $\gamma_o$, respectively.

Since this is such a simple model, we can compute the Fisher matrix analytically under certain assumptions.
We first compute the Fisher matrix with elements $\{M_3,\Omega_o,\sin\iota_J\}$. 
After computing the derivatives of $\tilde{h}$, we expand in $e_o\ll 1$ and consider only the secular effect of the trigonometric functions in $\xi_r$ and $\xi_\varphi$, as the contributions from oscillating terms will be minimal after integration over $f$. We substitute $t\propto \mathcal{M}^{-5/3}f^{-8/3}$ (cf. Eq. \eqref{eq:t vs f}) and after integrating over $f$ with a flat PSD, we find that the resulting matrix is invertible only if we have some pericenter precession $\epsilon \ne 0$. The fractional error in mass then scales like
\begin{equation}
    \frac{\Delta M_3}{M_3}\propto\frac{1}{\epsilon}\Bigr(\frac{a_o}{M_3}\Bigr)^{3/2}.
\end{equation}

This result shows that pericenter precession will produce contours of constant $\Delta M_3/M_3$ which scale with $(a_o/M_3)^{3/2}$, which is in good agreement with the lines in Fig. \ref{fig:BDEC_measure_m3_contours}, for example.
These lines do not appear when $e_o=0$, however, because the difference in the radial and azimuthal frequencies does not produce any change in the BBH's actual orbit when that orbit is circular.

We also expanded the Fisher matrix dimensions to include $\{M_3,\Omega_o,\sin\iota_J,\xi_{r,0},\xi_{\varphi,0},t_c\}$ and found identical scaling in $\Delta M_3/M_3$.

In contrast, let us examine the behavior of dS precession. We will now show how mass and orbital frequency can be independently measured by the the Thomas phase $\Phi_T$ and the polarization phase $\Phi_P$.
As discussed in \cite{Apostolatos:1994mx}, the Thomas phase for a circular orbit is approximately
\footnote{This comes from Eq.~(65) of \cite{Apostolatos:1994mx}. There are multiple expressions for the average rate depending on the sign and dot products of $(\hat L, \hat N)$, which only differ by factors of of order unity.}
\begin{equation}
    \Phi_T \sim 2 \pi (1 + \cos \lambda_L) \Omega_\text{dS} t \, ,
\end{equation}
where $\Omega_\text{dS} = \frac{3}{2} \frac{M_3 \Omega_o}{a_o} \approx \frac{3}{2} M_3^{2/3} \Omega_o^{5/3}$.

The Thomas phase breaks the degeneracy between $M_3$, $\Omega_o$, and $\sin\iota_J$.
If we consider a waveform with perfectly measured $\Omega_o$, with a simple waveform $h \propto e^{i \Phi_T}$ one can show the measurement scales as
\begin{multline}
\label{eq: dOmegadS/dlogM3}
    \frac{\Delta M_3}{M_3} \approx  \sqrt{\frac{1}{\Gamma_{\log M_3 \log M_3} }} \propto \frac{1}{\partial_{\log M_3} \Omega_\text{dS}} \\ 
    = \frac{1}{M_3^{2/3} \Omega_o^2} \propto \frac{a_o^{5/2} }{M_3^{3/2}}
\end{multline}
For contour plots such as Fig. \ref{fig:BDEC_measure_m3_contours}, this result matches the power law for lines of constant $\Delta M_3/M_3$ found in the region where de Sitter precession is most rapid. 

The polarization phase breaks the degeneracy between $M_3$ and $\Omega_o$ in the same manner as the Thomas phase. The power law can be seen by considering a waveform $h \propto e^{i \Phi_P}$. The polarization phase is defined as

\begin{equation}
\label{eq:tan phiP}
    \Phi_P(t)=- \arctan \left[ \frac{A_\times(t)F_\times(t)}{A_+(t)F_+(t)}\right].
\end{equation}
In the context of the Fisher matrix, it is useful to compute the derivative of this phase,
\begin{equation}
    \partial_\theta \Phi_P= -\frac{1}{(1+\tan^2\Phi_P)} \partial_\theta\Bigr(\frac{A_\times(t)F_\times(t)}{A_+(t)F_+(t)}\Bigr).
\end{equation}

The fraction of amplitude factors and antenna patterns depends on the angles $\theta_S,\phi_S,\theta_J,\phi_J,\lambda_L,\text{ and }\alpha_0$, as well as the integrated de Sitter precession rate, all of which appear exclusively in trigonometric functions (cf.~Eqs.~\eqref{eq: alpha(t)}, \eqref{eq: aplus}, \eqref{eq: across}, \eqref{eqn:detframe1}, \eqref{eqn:detframe2}, and \eqref{eq:psiS}). We assume the instantaneous de Sitter precession rate does not vary significantly, assigning the rate to its secular value $\Omega_{dS}\approx\frac32\frac{M_3\Omega_o}{a_o}$. 

Taking $\theta=\log M_3$, then, the only dependence on $M_3$ in this fraction is through $\Omega_{dS}$. Since $\Omega_{dS}$ appears only in the argument of sines and cosines, we expect that after applying the chain rule $\partial_{\log M_3}\rightarrow (\partial \Omega_{dS}/\partial \log M_3)\partial_{\Omega_{dS}}$, the magnitude of $\partial_{\log M_3}\Phi_P$ is primarily influenced by $\partial \Omega_{dS}/\partial \log M_3$. This expression evaluates to (cf. Eq.~\eqref{eq: dOmegadS/dlogM3})
\begin{equation}
    \frac{\partial \Omega_{dS}}{\partial \log M_3}=\frac{M_3^{3/2}}{a_o^{5/2}},
\end{equation}
Thus, assuming $\Omega_o$ is known, the Fisher matrix will scale like
\begin{equation}
    \Gamma_{\log M_3 \log M_3} = \int df \left(\partial_{\log M_3} \Phi_P(f)\right)^2 \propto (\partial_{\log M_3} \Omega_\text{dS})^2.
\end{equation}
Therefore, considering both $\Phi_T$ and $\Phi_P$, de Sitter precession produces an uncertainty in $\log M_3$ that scales roughly as $a_o^{5/2}/M_3^{3/2}$, following lines of constant $\Omega_{dS}$.

It may seem counter-intuitive that two processes with identically-scaling rates (i.e., pericenter precession and de Sitter precession) produce different power laws in the shape of the contour lines. However, even though the rate of pericenter precession $\Omega_\text{peri prec} \propto M_3^{2/3} \Omega_o^{5/3}$ occurs at the same PN order as de Sitter precession $\Omega_\text{dS} \sim M_3^{2/3} \Omega_o^{5/3}$, there is an additional factor of $a_o$ in the pericenter precession contribution due to the Doppler shift term being proportional to $r$. This causes the power law dependence for the measurement accuracy of $M_3$ to scale differently by a factor of $a_o$.

\end{document}